\documentclass[10pt]{article}
\usepackage[top=0.85in,footskip=0.75in]{geometry}
\usepackage{amsmath,amssymb}
\usepackage{changepage}
\usepackage[utf8x]{inputenc}
\usepackage{textcomp,marvosym}
\usepackage{cite}
\usepackage{nameref}
\usepackage{microtype}
\DisableLigatures[f]{encoding = *, family = * }
\usepackage[table]{xcolor}
\usepackage{array}
\usepackage{fixltx2e}
\newcolumntype{+}{!{\vrule width 2pt}}
\newlength\savedwidth

\newcommand\thickhline{\noalign{\global\savedwidth\arrayrulewidth\global\arrayrulewidth 2pt}%
\hline
\noalign{\global\arrayrulewidth\savedwidth}}
\usepackage{setspace} 
\raggedright
\usepackage[aboveskip=1pt,labelfont=bf,labelsep=period,justification=raggedright,singlelinecheck=off]{caption}

\makeatletter
\renewcommand{\@biblabel}[1]{\quad#1.}
\makeatother
\date{}

\usepackage{lastpage,fancyhdr,graphicx}
\usepackage{epstopdf}
\begin{document}
\vspace*{0.2in}

\begin{flushleft}
{\Large
\textbf\newline{Blood Vessel Tortuosity Selects against Evolution of Agressive Tumor Cells in Confined Tissue Environments: a Modeling Approach} 
}
\newline
\\
Andr\'{a}s Szab\'{o}\textsuperscript{1*\textcurrency} and 
Roeland M.~H.~Merks\textsuperscript{1,2}
\\
\bigskip
\textbf{1} Life Sciences Group, Centrum Wiskunde \& Informatica, Amsterdam, The Netherlands
\textbf{2} Mathematical Institute, Leiden University, Leiden, The Netherlands
\\
\bigskip

\textcurrency Current Address: Department of Cell and Developmental Biology, UCL, London, United Kingdom



* A.Szabo@ucl.ac.uk

\end{flushleft}
\section*{Abstract}
Cancer is a disease of cellular regulation, often initiated by genetic mutation within cells, and leading to a heterogeneous cell population within tissues. In the competition for nutrients and growth space within the tumors the phenotype of each cell determines its success. Selection in this process is imposed by both the microenvironment (neighboring cells, extracellular matrix, and diffusing substances), and the whole of the organism through for example the blood supply. In this view, the development of tumor cells is in close interaction with their increasingly changing environment: the more cells can change, the more their environment will change. Furthermore, instabilities are also introduced on the organism level: blood supply can be blocked by increased tissue pressure or the tortuosity of the tumor-neovascular vessels. This coupling between cell, microenvironment, and organism results in behavior that is hard to predict. Here we introduce a cell-based computational model to study the effect of blood flow obstruction on the micro-evolution of cells within a cancerous tissue. We demonstrate that stages of tumor development emerge naturally, without the need for sequential mutation of specific genes. Secondly, we show that instabilities in blood supply can impact the overall development of tumors and lead to the extinction of the dominant aggressive phenotype, showing a clear distinction between the fitness at the cell level and survival of the population. This provides new insights into potential side effects of recent tumor vasculature renormalization approaches.

\section*{Author Summary}
Multicellular organisms control their cells to facilitate higher level function of the whole organism. In tumors, this control is lost and cells are allowed to enhance their fitness by, for example, increased proliferation. Tumor cells continue to change their behavior through accumulating mutations, leading to a complex and highly heterogeneous structure. Several computational studies have investigated the emergent structures of such mutating group of cells and led to the recognition that cellular heterogeneity within tumors is essential to explain the observed morphologies. Most of these studies have considered a limited number of possible cell phenotypes, an isolated tumor cell population, unlimited growth space around the tumor, often with an unexhaustable source of nutrients. Here we introduce a modelling approach that takes into account the limited growth space around the tumor, localized nutrient sources, cellular metabolism, and mutation in a continuous phenotype space. The model reproduces the Warburg effect due to the limited nutrient supply leading to an irreversible switch in cellular metabolism, and, consistently with previous models, exhibits stages of development together with a natural selection for a rapidly growing phenotype. This phenotype locally emerges in stable environments, but when nutrient supply becomes erratic, these show less resilience and are outcompeted by slow growers.


\section*{Introduction}
Cancer is a disease of multicellular regulation, in which malfunctioning cells can break free of homeostatic regulations imposed by the host environment \cite{Hanahan2011}. One of the main characteristics of cancer is the increased proliferation and mutation of cancerous cells due to malfunctioning control of growth and proliferation \cite{Hanahan2011}. As these behavioral changes typically originate from mutations in the cells' genetic material, excessively proliferating cells accumulate further alterations, leading to a possible amplification of malignancies. Traditional studies of altered cell traits primarily focus on genetic mutations, but neglect the multicellular nature and genetic variety of tumors. 

Tumor heterogeneity has been demonstrated experimentally and is an active field of research \cite{Gerlinger2012, Sottoriva2013, Yuan2012a}. Neutral mutations may accumulate and contribute to intratumor heterogeneity \cite{Williams2016}. An intermediate level of heterogeneity is correlated with low survival probability \cite{Andor2015}. Heterogeneity may even promote the collapse of tumor development by inducing a clone population that supports and enhances the growth of other clones in mice \cite{Marusyk2014}. Marusyk and colleagues \cite{Marusyk2014} claim that these supporter clones may be outcompeted by the more aggressive subpopulation, leading to the disappearance of the supporters and to the consequent collapse of the tumor. Heterogeneity questions the validity of previous whole--tumor analyses, as ``the most abundant cell type might not necessarily predict the properties of mixed populations" \cite{Marusyk2012}, and emphasizes the need for more detailed approach. 

Initial phases of tumor development are increasingly thought to give rise to a Darwinian process \cite{Hanahan2011, Little2010}, where individual cells compete for growth space and nutrients. Selection is imposed by the microenvironment, a highly complex entity spanning several ranges in size from the endocrine regulation of the whole body down to the extracellular matrix (ECM) and neighboring cells. During cancer development this environment is changed due to changes in the cellular component, and due to the tissue and organism level reactions to the tumor. 

Intrinsic coupling between neighboring cells forms the basis of the plasticity-reciprocity model \cite{Friedl2011}: as cells alter their behavior (plasticity), their contribution to the local environment changes through for example ECM remodelling, nutrient uptake, or adhesion molecule expression. In turn, the changed environment imposes an altered selection on the cells (reciprocity), creating a feedback between cells and their local microenvironment. This cascading change in behavior is reminiscent of the behavioral changes associated with stages of cancer development \cite{Foulds1958} that was later described by the accumulation of mutations through which cells become increasingly malignant \cite{Fearon1990}. Although a strict sequence of mutations was not found, a general pattern was observed in the majority of cases \cite{Fearon1990}, linking the macroscopic stages to the cell-level changes.

Computational modeling is an excellent tool for exploring, studying, and understanding such complex systems, because it provides complete control over assumptions and reveal the consequent behavior emerging from them. This allows the dissection of complex interactions and exploration of experimentally challenging cases. A variety of models have been applied to study cancer. 

Using a population level description, Basanta and co-workers studied how the combination of tumor treatments, p53 cancer vaccine and chemotherapy, can be optimized to yield the best results \cite{Basanta2012}. In a similar study, Sreemati Datta and coworkers \cite{Datta2013} model tumor development with the inclusion of evolving mutation rates and show that the balance between inducing driver mutations and mutation rates plays a key role in tumor growth: at high mutation rates, genetic instability may counter tumor progression. Combined with close experimental verification, Marusyk and colleagues \cite{Marusyk2014} used a similar model to suggest that interactions among clones may lead to an overall collapse of tumor development. This approach is able to incorporate evolutionary games to help cope with the development of treatment resistance. Such space-free models assume that all cells within the tumor are able to interact with all other cells directly, and are unable to deal with intra-tumor spatial heterogeneities. 

Studies incorporating this spatial aspect have mostly worked with cellular automata (CA) models. For example, Gerlee and colleagues were able to explain the `go or grow' hypothesis through the emergence of haptotaxis in their CA model \cite{Gerlee2009}. In a recent study, Waclaw and colleagues \cite{Waclaw2015} used a CA model to show that cell motility together with cell turnover may prevent intratumor heterogeneity. Of particular interest is the study of Anderson and colleagues \cite{Anderson2006}, where the evolution of a growing cell population and the effect of a heterogeneous environment is explored. They represent the tumor environment as a distribution of ECM molecules that together with oxygen serve as nutrient after degradation. Cell evolution is modeled in the phenotype by selecting a set of cellular parameters (matrix degradation rate, proliferation rate, etc.) from a number of predefined phenotypes upon cell division. The new phenotype was either selected randomly or according to a predefined sequence progressing towards more aggressive behavior. The authors found that cells evolved into a similar, aggressive phenotype when applying the random mutation scheme. Heterogeneity in the population was found to give rise to irregular tumor surface, whereas environmental heterogeneity (heterogeneity in the ECM distribution) reduced population heterogeneity and favored the most aggressive cell types. Low concentration of oxygen was also found to reduce population heterogeneity and promote invasive finger formation; applying two bursts of oxygen in these simulations led to the segregation of the mixed population. Similar models have been used to show emergent progression of phenotypes related to hypoxia, glycolysis, and acid-resistance, by including neuronal networks in cells \cite{Gerlee2008}, or angiogenesis by introducing blood vessels in hypoxic regions \cite{Robertson-Tessi2015}. Enderling and co-workers \cite{Enderling2009} introduced the cancer stem cell (CSC) hypothesis in a similar model by incorporating cells with unlimited proliferation potential (CSCs) and cells with limited proliferation potential. They found that cell death induced by tumor therapy could lead to a more aggressive, proliferative tumor, as the CSCs were no longer competing for space after the treatment. Using a combination model of phenotype evolution and the CSC hypothesis, Sottoriva and colleagues showed that the presence of CSCs in the model tumors led to more invasive tumor morphology \cite{Sottoriva2010}. The CA model further allowed exploring efficient drug delivery through the neovasculature \cite{McDougall2006, Stephanou2005}, the role of spatial arrangement of the vasculature in radiation therapy \cite{Scott2016} and even the effect of the different 2D representations of the 3D vasculature \cite{Grogan2016}. 

Using a more detailed model allows to explore the evolution of many other aspects of cell behavior, such as cell flexibility, cell adhesion, or cell shape. One such model used in tumor modeling is the cellular Potts model \cite{Graner1992, Szabo2013}. It has been used to describe, for example, the effect of nutrient limitation on tumor growth morphology \cite{Poplawski2009}, or to compare the emergence of distinct developmental stages in terms of morphology and growth in vascular versus avascular tumors \cite{Shirinifard2009}. Using a heterogeneous evolutionary model of CSCs, a recent study reported the emergence of spatial stratification of tumors based on the evolution of adhesion molecule expression \cite{Swat2015}. A further line of models explore the effect of oxygenation of tumors \cite{Powathil2012} in the light of chemotherapy \cite{Powathil2014}. An alternative way of introducing more detail is a spatial continuum representation of the cells. One example of this is the phase field method which has been used in combination with discrete modelling elements to investigate tumor growth morphology and its interaction with the vasculature \cite{Frieboes2010,Yan2016}.

Most of the above models applies a sharp distinction between the tumor and the healthy tissue. While the tumor is described in a cell-based detail, the environment is usually presented as a continuum. These models implicitly assume that stromal cells surrounding the tumor do not participate in the development of the tumor, and that these cells are fundamentally different from the cancerous tissue. The transition between the stromal and cancerous cell states is neglected. A recent exception from this is the work of Powathil and colleagues \cite{Powathil2016} where the effect of irradiation on a small set of ``healthy" bystander cells is investigated. As a result of irradiation and induced signals, stromal cells may apoptose or may be converted to the tumor cell type in the model. However, in this and almost all of the above models the phenotypes available for evolving cells are restricted to a small, discrete subset of all possibilities. A continuous range of cell states could reflect a more biological picture including more subtle, for example epigenomic, changes. Furthermore, the nutrients are typically modeled as a single component, however, one signature of cancerous cells is the reduced oxygen consumption and increased glucose uptake. For this a more detailed nutrient description is required. Another less explored point of interest is the change in the temporal behavior of the larger environment, the nutrient supply. Initial tumors experience a stable blood supply in healthy tissues, that presents them with a relatively constant environment. In contrast, angiogenic tumors at a later stage develop neovasculature that is tortuous and leaky, presenting a fluctuating, unstable environment for the cells \cite{Carmeliet2011, Weis2011}. How does this temporal variation affect the population behavior? Would the same aggressive phenotype dominate the population as in the stable environment, or would it fit less than the less aggressive and stable phenotype? 

\begin{figure}
\begin{centering} 
\includegraphics[width=1.00\textwidth]{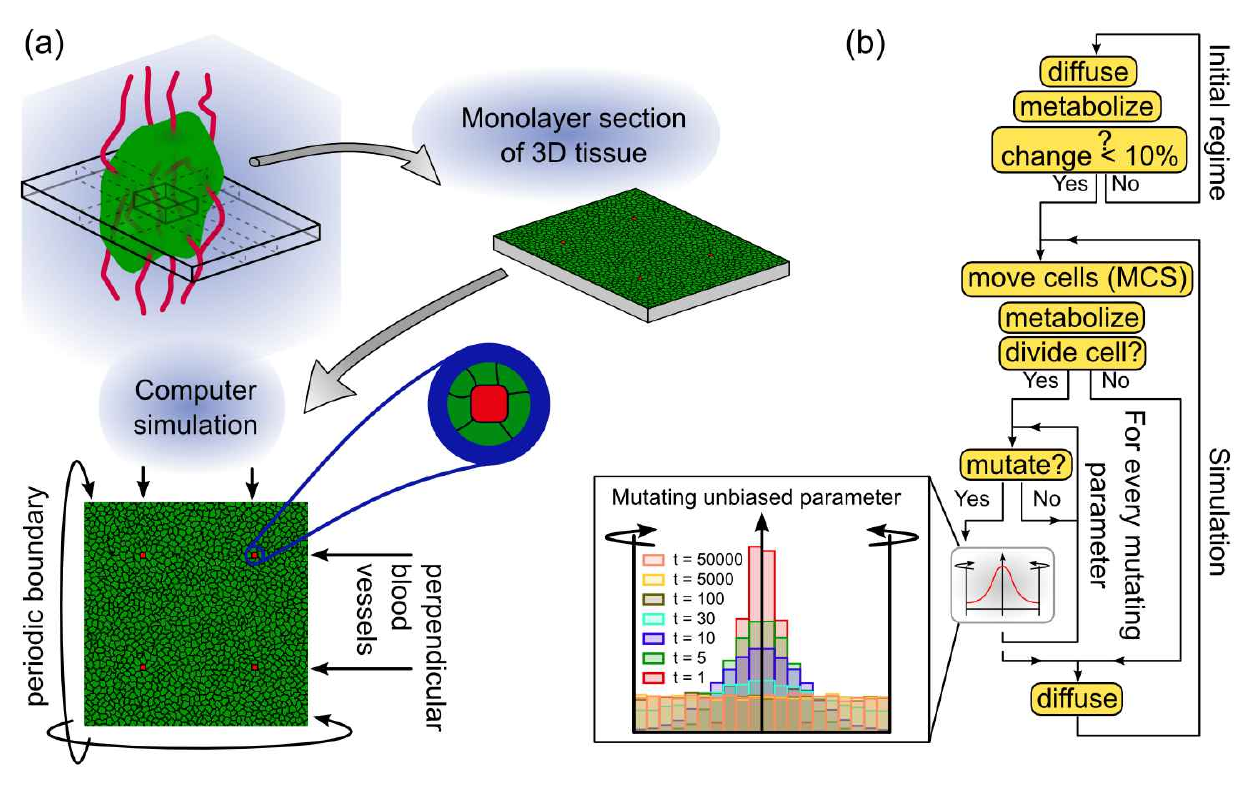}
\caption{\textbf{Illustration of model configurations and workflow of simulations.} (a) Cells are initialized as a monolayer with four endothelial cells (red) representing cross-section of blood vessels perpendicular to the plane of simulation. (b) Simulation workflow. In the initial regime nutrients are allowed to diffuse into the system to generate a steady-state distribution, during which cells metabolize without changing their size or phenotype. After the fields have stabilized, the simulation is started and is run in iterative cycles of Monte Carlo steps (MCS). After each MCS, each cell performs metabolic trafficking, and is probed for cell division. Upon division, a potential mutation event is triggered in the daughter cells. Inset: unbiased parameters from $10^4$ cells after $t=1, 5, 10,..., 5\times10^4$ number of mutations are distributed uniformly in the algorithm.} 
\label{fig_modelSetup} 
\end{centering} 
\end{figure}

Here we present a cellular Potts model of a closely packed, mutating cell population representing an epithelial tissue within an organism (Fig.~\ref{fig_modelSetup}). Mutation in the model allows cellular behaviors to vary continuously in a wide range of phenotype space, therefore evolution is governed by selection emerging naturally from the limitations of space, glucose, and oxygen. Cellular metabolism is modeled as a mixture of aerobic glycolysis and respiration. Small scale changes in the microenvironment are represented by the changes of the immediate cellular neighborhood, both by mutations and cell rearrangements. We model large scale environmental fluctuations through the fluctuating activity of spatially fixed nutrient sources, which represent the cross-section of blood vessels.

Constructed in this way our model includes both the plasticity-reciprocity model of Friedl and Alexander \cite{Friedl2011}, as well as the large scale fluctuations imposed by the host. In the following sections we analyse the behavior of the model with and without mutating cells. We show that our model reproduces the Warburg shift, and exhibits stages of development similar to those observed in previous studies (such as \cite{Anderson2006, Gerlee2008, Robertson-Tessi2015}). In our model cells undergo clonal expansion, hypoxia, followed by starvation, with the development of segregated populations around blood vessels. The spatial differentiation of cell populations is somewhat similar to the spatial diversity in real tumors as described by Alfarouk et al.~\cite{Alfarouk2012}. Whereas Alfarouk and colleagues describe two main habitat zones concentrically surrounding the blood vessel, we observe only one of the zones with high proliferation rates and a robust cellular outflow from near the nutrient source. Finally, our results indicate that the dominant aggressive phenotype is more sensitive to fluctuations in the environment than the ones maintaining a stable phenotype without mutation.

\section*{Results}

\subsection*{Cellular Potts model of a homeostatic tissue}

To investigate the above questions, we model a monolayer of cells using a modified cellular Potts model (CPM) based on the CompuCell3D implementation \cite{Swat2012} which can be obtained from http://www.compucell3D.org. Customized code for the simulations and example parameter and initial condition files can be found in \nameref{S1-File}. In the following we give an overview of the model; for more detail see the Methods section. 

Cells in the CPM are represented as confluent domains on a lattice on which an integer $\sigma(\vec{x})$ at every position $\vec{x}$ indicates which cell is occupying the location $\vec{x}$; cell-free areas are designated by $\sigma(\vec{x})=0$. Cell movement results from a series of elementary steps in which an attempt is made to copy $\sigma(\vec{x})$ at a randomly selected location $\vec{x}$ to one of its randomly selected neighboring location $\vec{x'}$. This attempt is accepted with a probability based on a Hamiltonian goal function $H$ that defines cell dynamics (Eqs.~\ref{probability}, \ref{Hamiltonian}). $H$ is usually defined such that cells maintain a controlled size, perform amoeboid-like cell movement, and may exhibit adhesion or contact-repulsion. A time step in the model is defined as the Monte Carlo Step (MCS) consisting of $N$ elementary steps where $N$ is the total number of lattice sites in the model. In our model we apply the usual calibration by relating 1 MCS to 1 minute real time, and 1 lattice site to 2 $\mu$m. Diffusion of soluble substances are simulated on a lattice identical to the cellular-lattice. This calibration relates the simulated tissue area to $400\mu m \times 400\mu m$, and diffusion coefficients of simulated nutrients to realistic values (glucose and lactate: $D_g=10^{-9}m^2/s$; oxygen: $D_{O_2}=D_l=10^{-11}m^2/s$) \cite{Jiang2005}. 

We implemented a metabolism whereby cells consume glucose and oxygen from their environment and use a mixture of lactic acid fermentation and cellular respiration. Cells metabolize oxygen and glucose to generate an abstract cellular energy that is used for their maintenance and growth. The amount of energy required for cells is controlled by the expression levels of glucose transporters in the cell membrane which is determined by an intracellular growth signal parameter ($N_0(i,t)$ for cell $i$ at time $t$). The mode of metabolism is determined by an internal hypoxia inducible factor ($h(i,t)$ for cell $i$ at time $t$), that is controlled through the oxygen levels inside and around the cell and the amount of intracellular reactive oxygen species (ROS)(Eq.~\ref{eq:hif}). This factor determines the ratio of respiratory and fermentative modes of cellular metabolism. In our model the speed of energy production, and hence cell growth, is independent of the mode of metabolism in order to avoid a selection bias towards the faster metabolic mode. 

If a cell reaches a pre-defined doubling size it divides, hence cell cycle time is determined by cell metabolism. Cell death may occur either due to starvation or age. If a cell generates less energy from metabolism than is required for maintenance, it converts the necessary amount of its cell mass into energy (catabolism). Once the cell mass is exhausted, the cell is considered dead and is taken out of the simulation. Cells are also killed in the simulation with a 0.1\% probability after each MCS to maintain cell turnover. 

Glucose and oxygen are supplied by a separate set of designated immobilized cells that play the role of blood capillaries (Fig.~\ref{fig_modelSetup}a). To allow the temporal control of nutrient supply, capillaries can be in an active or a blocked state. Nutrient levels are kept at a fixed concentration in the blood stream, that is inside the capillaries, when the capillaries are active. When a capillary is blocked, nutrient levels are kept at zero within the capillaries. The activity of the vessels are changed with a probability (nutrient switching probability) after each MCS. A high switching probability leads to a stable supply while a low switching probability results in an inconsistent supply, mimicking blocked or tortuous vessels without affecting the average activity time of the vessels. Lactate produced by cells as a waste from lactic acid fermentation is cleared out of the system by active capillary cells. 

For more detail on the model see the Methods section. 

\begin{figure}
\begin{centering} 
\includegraphics[width=1.00\textwidth]{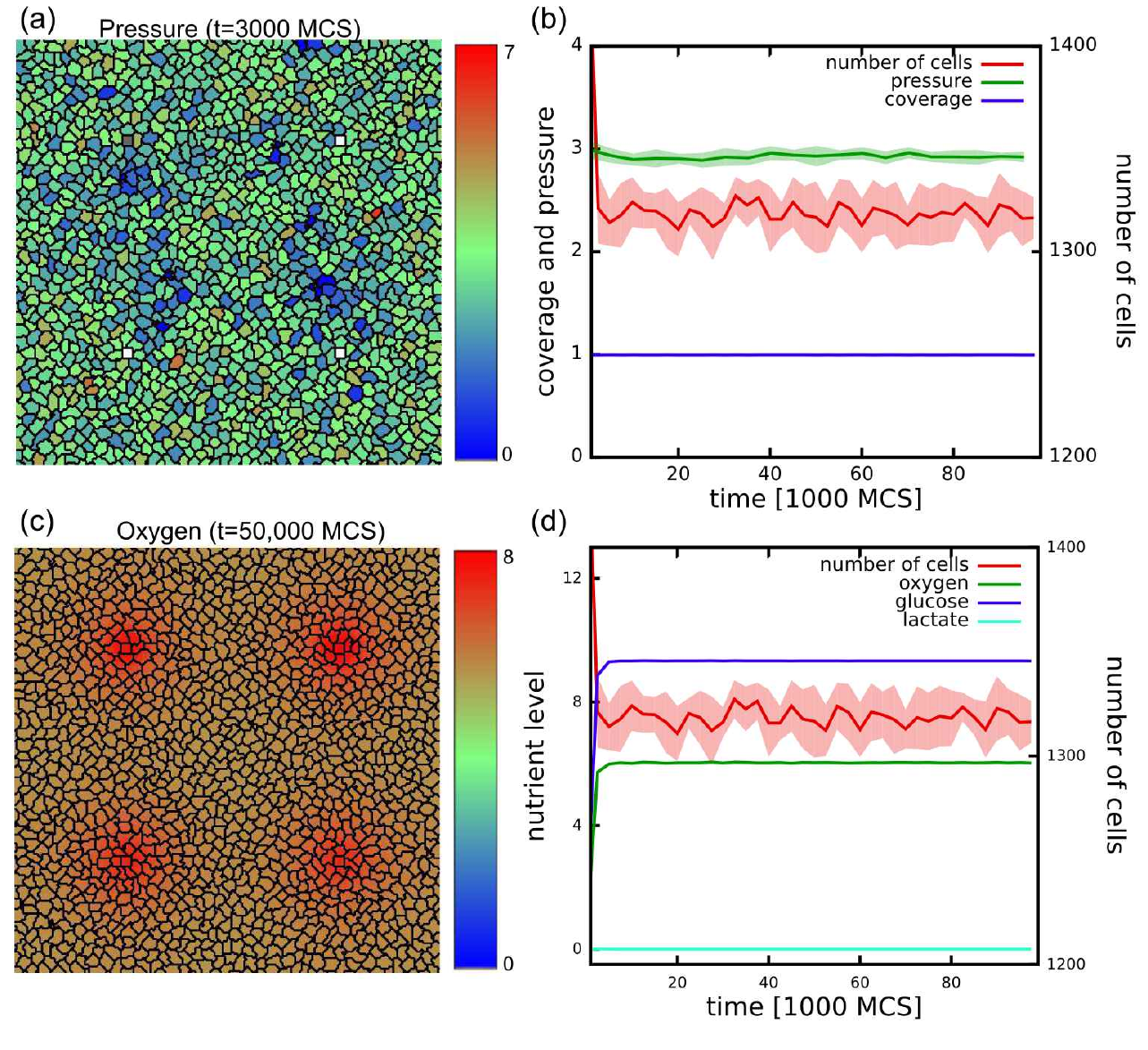}
\caption{\textbf{Simulations without mutations and with a stable nutrient source result in a stable tissue.} (a) Cell configuration at time t = 3000 MCS, showing pressure as color. Note the lack of any specific pattern in the distribution of intracellular pressure. (b) The number of cells, and intracellular pressure stay relatively constant in the course of the simulations. Cells cover approximately the whole of the tissue (coverage). Data averaged from 10 independent simulation runs. (c) Nutrient distribution in the system happens via diffusion from active blood vessels. The color codes the oxygen concentration in the system at t = 50,000 MCS. (d) Nutrient levels stay constant in the system with abundant glucose, somewhat less oxygen, and no lactate. The absence of lactate indicates that cells in the population remain respiratory. } 
\label{fig_stableStages} 
\end{centering} 
\end{figure}

We start by first verifying that the model is capable of simulating a sustainable homeostatic tisse. Therefore we studied the behavior of a healthy tissue in the absence of mutation and stable nutrient supplies. Fig.~\ref{fig_stableStages}a shows the model setup. We consider a two-dimensional square lattice, corresponding with a slice of tissue of 400 $\mu$m $\times$ 400 $\mu$m (40 $\times$ 40 cells, 200 $\times$ 200 lattice sites), containing four blood vessels arranged in a square formation (Fig.\ref{fig_modelSetup}a). To achieve constant nutrient supply, the probability of a vessel to be blocked or unblocked in every MCS is 0.5. In this case each vessel switches between active (depicted in white) and blocked (depicted in gray) states rapidly. As the nutrients diffuse away from the source, this rapid switching results in a continuous supply of nutrients (Fig.~\ref{fig_stableStages}c). The color of tissue cells in Fig.~\ref{fig_stableStages}a indicates the intracellular pressure, defined as the difference between target volume (biomass) and actual volume of the cell ($\pi(i,t)=V^T(i,t) - V(i,t)$). This measure differs from the extracellular pressure as it includes any contribution of the contractile actin cortex surrounding biological cells. Note that the pressure within the population is distributed without any specific pattern. 

Fig.~\ref{fig_stableStages}b shows the dynamics of the tissue for $10^5$ MCS corresponding to approximately 70 days. The number of cells fluctuates around a constant value throughout the simulation (Fig.~\ref{fig_stableStages}b), and is sufficient to cover the whole system: the ratio of the cell-covered region over the total simulation area is close to one (Fig.~\ref{fig_stableStages}b). As the average intracellular pressure does not increase over the course of the simulation (Fig.~\ref{fig_stableStages}b), cell growth and proliferation are kept in balance with the basal metabolism and the constant cell turnover. Excessive growth is prevented by the lack of growth space through a negative feedback between the intracellular pressure and cell growth (Eq.~\ref{eq:cellGrowth}). Nutrient levels remain constant during the simulations and cells remain respiratory as indicated by the absence of lactate (Fig.~\ref{fig_stableStages}d).

\subsection*{Model of tissue micro-evolution}

Next we investigated how a cancerous tissue would behave in our model. Cancerous tissues are characterized by large number of mutations, chromosomal rearrangements and changes in gene expression level, all of which may result in phenotypic changes of the cells \cite{Stephens2011, Hasty2014, Hanahan2011, Huang2006}. To mimic such changes, we allowed the set of 10 assigned phenotypic properties of cancerous cells to change upon division (Fig.~\ref{fig_modelSetup}b), including the division volume or adhesion parameters (see Table~\ref{table} and Methods). After cell division, the daughter cells inherit the phenotypes of their parents with some small mutations. Every parameter is allowed to change with a fixed probability (mutation rate $\mu_p$ for parameter $p$) and independently of one another. The change in parameter $p$ is drawn from a normally distributed random variable with a standard deviation of $\sigma_p$, that is:  $p'=p+\mathcal{N}(0,\sigma_p)$. The parameters are allowed to change freely within a pre-defined range (Table\ref{table}) with reflective boundary conditions. This allows an unbiased parameter to uniformly explore the available range (Fig.~\ref{fig_modelSetup}b inset). 

\begin{table}[ht]
\begin{adjustwidth}{0in}{0in} 
\centering
\caption{
{\bf List and values of mutating parameters in the model.}}
\begin{tabular}{l l c c c c}
\hline
\hline
{\bf Parameter} & {\bf Notation} & {\bf Initial} & {\bf Step} & {\bf Minimum} & {\bf Maximum} \\
{\bf name}      &          & {\bf value}   &{\bf size} & {\bf value}   & {\bf value}   \\
\thickhline
Incompressibilty & $\lambda_{v}$ & 25 & 0.2 & 0 & 20    \\
Division volume & $V_D$ & 50 & 1 & 10 & 200  \\
Cell adhesion & $\rho_{\text{CAM}}(\text{cell})$& 99 & 0.1 & 0 & 100 \\
Matrix adhesion & $\rho_{\text{MAM}}(\text{cell})$ & 49.5 & 0.1 & 0 & 50 \\
Glucose chemotaxis & $\chi_{g}$ & 0 & 0.1 & -20 & 20 \\ 
Oxygen chemotaxis & $\chi_{O_2}$& 0 & 0.1 & -20 & 20 \\
Lactate chemotaxis & $\chi_{l}$& 0 & 0.1 & -20 & 20 \\
Growth (see Eq.~\ref{eq:nglut}) & $N_0$ & 250 & 25 & 0 & 2500 \\
HIF threshold & $\kappa_h$ & 0.1 & 10$^{-3}$ & 10$^{-4}$ & 10 \\
ROS threshold & $\kappa_\zeta$ & 0.5 & 0.01 & 10$^{-4}$ & 100 \\
\end{tabular}
\label{table}
\end{adjustwidth}
\end{table}

\begin{figure}
\begin{centering} 
\includegraphics[width=0.55\textwidth]{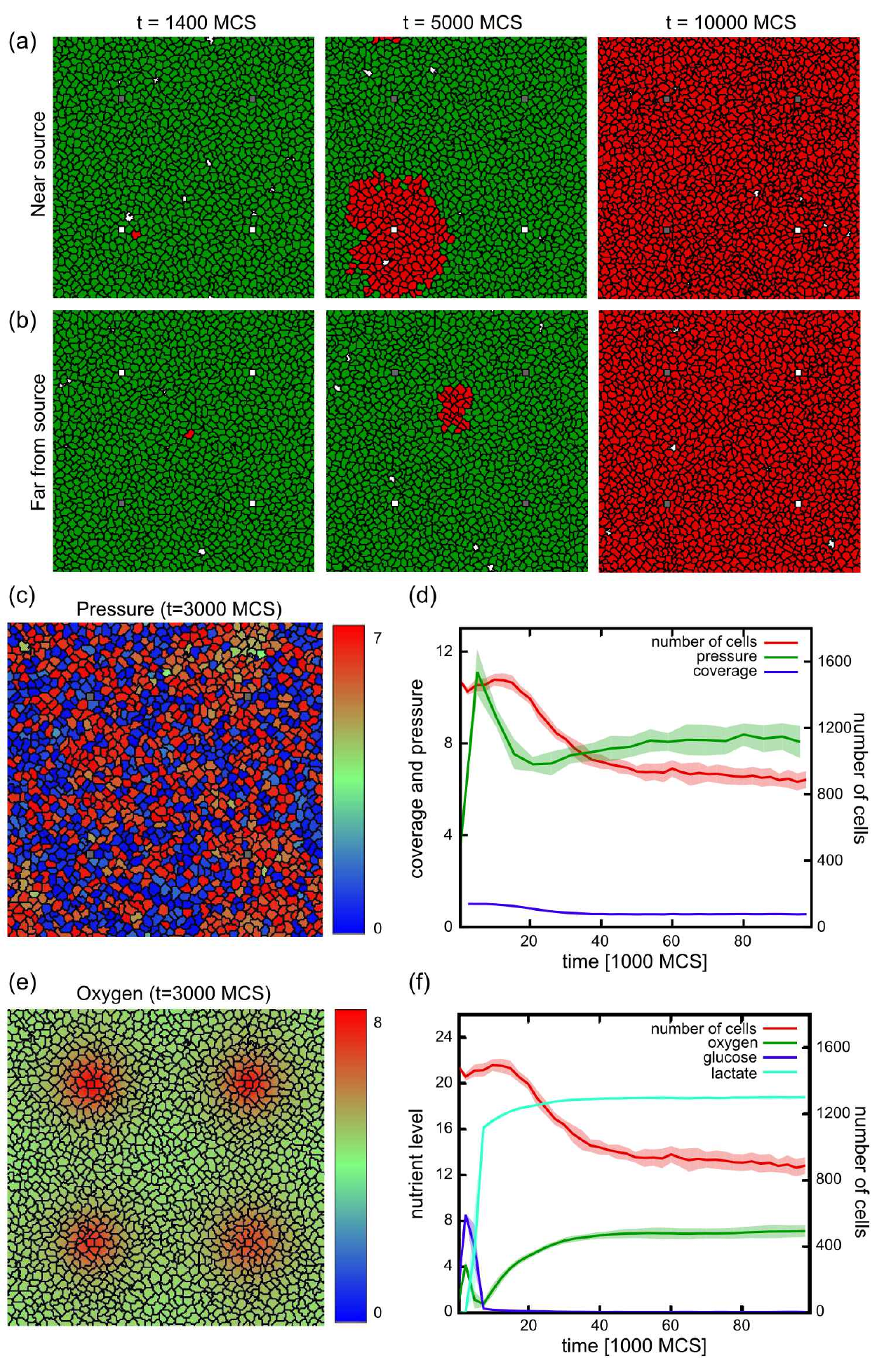}
\caption{\textbf{Simulations with mutating cells and stable vasculature.} Mutation rate = 10\%. (a-b) Simualtions initiated with non-mutating cells (green) and a single mutating cell (red) near (a) or far from (b) the nutrient source. The frames are from simulations where the mutating cell persisted and took over the population. (c-f) Results from simulations initiated with mutating cells. (c) Cell configuration at time t = 3000 MCS, showing intracellular pressure as the color code. Note the fluctuations in pressure distribution compared to the non-mutating case in Fig.~\ref{fig_stableStages}. (d) Population dynamics of a mutating population as a function of simulation time, showing the number of cells, intracellular pressure and tissue coverage. The number of cells initially increases and after a peak drops to almost to half of the peak value, where it stabilizes. The intracellular pressure follows the initial increase in cell numbers. During the cell number drop the pressure is moderated to almost half, after which it is increasing to reach a plateau. The tissue coverage drops from the initial full coverage to approximately 50\%. (e) Oxygen concentrations in the system at t = 3000 MCS. Nutrients are depleted further from the blood vessels, eventually resulting in a shortage of oxygen in distant regions. (f) Nutrient levels during the simulation reveal the depletion of glucose and oxygen during the expansion phase. A simultaneous increase in lactate levels shows the appearance of fermentors. When nutrients are depleted, the population size declines. With the decrease of respiration, oxygen levels return to the level observed in the non-mutating populations (Fig.~\ref{fig_stableStages}d). Plots show average of 10 independent simulations. } 
\label{fig_mutStages} 
\end{centering} 
\end{figure}

To determine how a single mutating cell would perturb the homeostasis of the above tissue, we inserted a cell with a mutation potential either near (Fig~\ref{fig_mutStages}a, red cell) or far from (Fig~\ref{fig_mutStages}b) the nutrient source. The single mutating cell persisted in 28 and 26 simulations out of 100 for the two different initiation positions. When the cell persisted, it expanded within the first 5000 MCSs and eventually colonised the population completely (Fig~\ref{fig_mutStages}a, b).

In order to study the internal dynamics of the tumor, we will only focus on the case where the mutating cell persists and has colonized the population; therefore we will initiate our simulations with populations where all cells are allowed to mutate. Fig.~\ref{fig_mutStages}c-f shows the behavior of the model with mutating cell populations, with 10\% mutation rate. In comparison with the non-mutating populations, the intracellular pressure is higher and exhibits a wider variation across neighbors (Fig.~\ref{fig_mutStages}c). This shows that cells overcome the initial growth control mechanism. The number of cells initially increases, reaches a peak, and then declines to approximately half of the peak value, well below the initial numbers, where the population size stabilizes (Fig.~ \ref{fig_mutStages}d). These changes in cell numbers are followed by the intracellular pressure as well: in the expansion phase the pressure increases, but before the peak in cell number it sharply declines. After the population size settles to a lower value, the pressure settles to an approximately constant value. The full coverage of the tissue drops to approximately half coverage, showing that the cancerous cells cannot maintain a complete monolayer. Nutrients are depleted further from the blood vessels, resulting in a shortage of oxygen in the distant regions (Fig.~\ref{fig_mutStages}e), shortly followed by the depletion of glucose (Fig.~\ref{fig_mutStages}f). The depletion of oxygen triggers the cells to switch to fermentation, resulting in an increase in lactate. This switch accelerates the depletion of glucose, causing the decline in population size. After the population size is reduced, oxygen levels return to the same level as in the non-mutating populations, while cells still rely on fermentation as can be seen from the maintained lactate levels (Fig.~\ref{fig_mutStages}f).

\subsection*{Emergent stages of development}

\begin{figure}
\begin{centering} 
\includegraphics[width=0.80\textwidth]{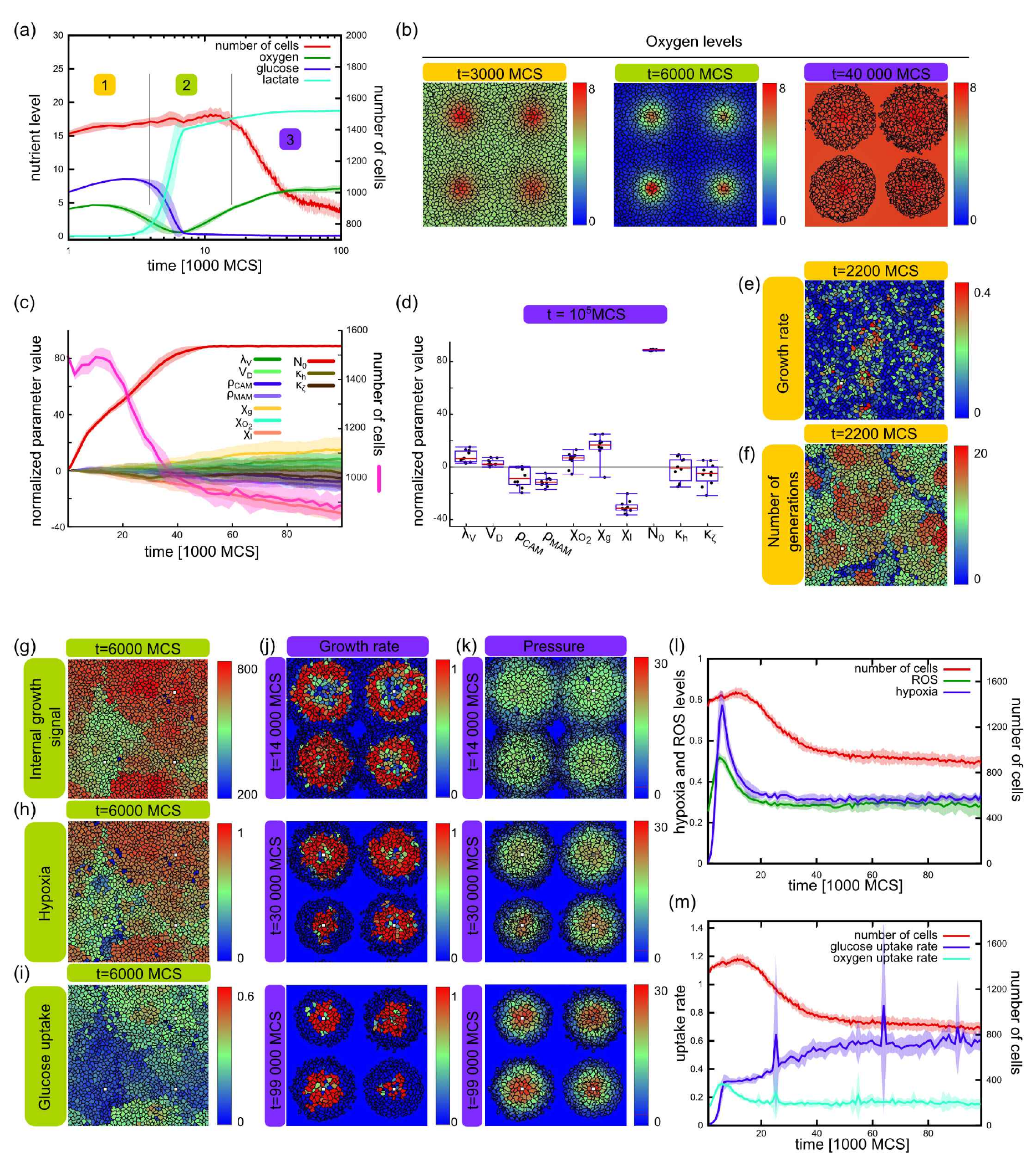}
\caption{\textbf{Stages of development in simulations with mutating cells and stable vasculature.} Mutation rate = 10\%. (a) Number of cells and nutrient levels as in Fig.~\ref{fig_mutStages}f, using a logarithmic time-scale with the three stages indicated with numbers. (b) Cell configurations of each stage with oxygen concentrations color coded. (c) Population size (pink) and averages of all 10 evolving parameters scaled with their respective step sizes $\sigma_p$ and shifted to their initial values. Intracellular growth signal: $N_0$. (d) Normalized parameter values at the end of the simulation runs ($t=10^5$ MCS). (e-f) Stage 1: expansion. Configuration of cells from a simulation showing the instantaneous growth rate (e) defined as the increase in target volume in the current MCS, and generation age (f) at t = 2200 MCS. Patches of high growth appearing independently from the localization of sources. (g-i) Stage 2: hypoxia. Configuration of cells from a simulation showing the intracellular growth signal $N_0$ (g), level of hypoxia (h), and glucose uptake (i) at t = 6000 MCS. (j-k) Stage 3: starvation. Configurations of cells from a simulation showing growth rates (j) and intracellular pressure (k) at late stages of development. (l-m) Evolution of cellular metabolism showing intracellular hypoxia and ROS (l) and nutrient uptake (m). Linegraphs show average of 10 independent simulations with standard deviation, boxplots show median with interquartile range and minimum / maximum values from 10 independent simulations.} 
\label{fig_progression} 
\end{centering} 
\end{figure}

Previous studies have reported distinct stages of development including hypoxia, glycolysis, or acid-resistance \cite{Anderson2006, Gerlee2008, Robertson-Tessi2015}. However, in these studies the evolution occurred in isolation from the stromal tissues and vasculature either using a limited set of phenotypic behaviors potentially constraining the degree of freedom of the evolutionary trajectories or with immobile cells. Therefore, we first asked if stages of tumor progression also occurred in our less constrained model. Fig.~\ref{fig_progression} shows the behavior of our model with the nutrient concentrations and cell numbers averaged from 10 independent simulations with stable vasculature (switching probability = 0.5) and 10\% mutation rate. Based on this, we identified distinct stages in our model: expansion (1), hypoxia (2), starvation (3).  Insets show cell configurations characteristic of the three stages, color scale on the insets indicates the oxygen concentrations. These stages emerge as a result of an interplay between the cells and their environment, as in the proposed plasticity-reciprocity hypothesis of Friedl and Alexander \cite{Friedl2011}. Stages in our model relate to: 1. Conditioning the environment; 2. A reaction to the environmental change in the behavior of cells (new phenotypes emerging, old phenotypes disappearing); 3. New environment created by the new population. Remarkably, this shows that despite the much larger number and freedom of mutating parameters in our model, we still find the same phenomena of emergent stages. 

\begin{figure}
\begin{centering}
\includegraphics[width=1.00\textwidth]{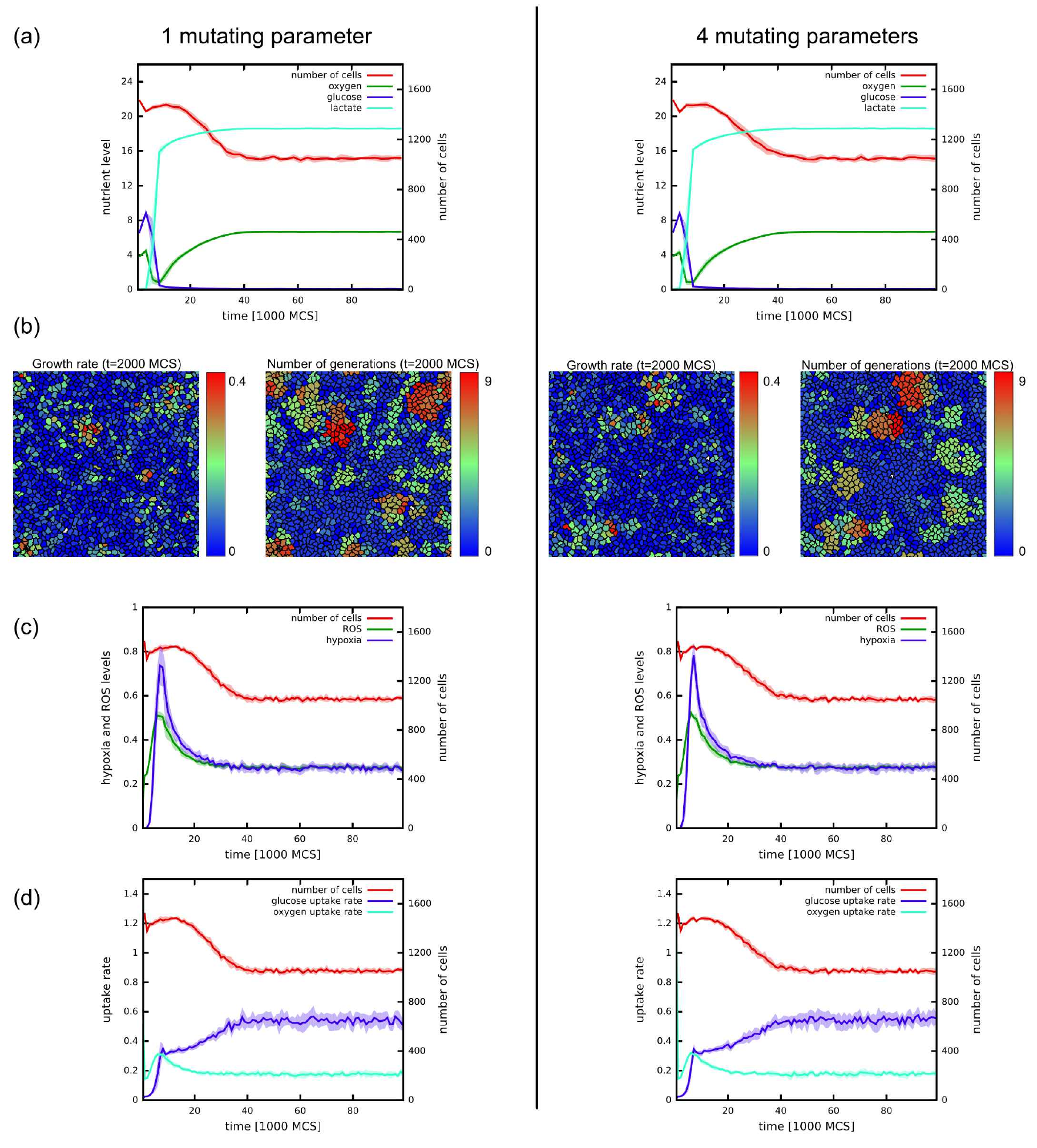}
\caption*{{\bf S1 Fig. Phenotypic behavior of simplified models with one ($N_0$) or four ($N_0, \chi_g, \chi_{O_2}, \chi_l$) mutable cellular parameters is similar to behavior of full model.} (a) Nutrient levels and number of cells in the simplified models; compare with Fig.~\ref{fig_mutStages}d). (b) Patches of high growth similar to those in the full model shown by the instantaneous growth rate and generation number of cells as in Fig.~\ref{fig_progression}e, f. (c) Hypoxia and ROS in the simplified models; compare with Fig.~\ref{fig_progression}l. (d) Oxygen and glucose consumption in the simplified models; compare with Fig.~\ref{fig_progression}m. Population averages from 10 independentent simulation runs with standard deviation accross simulations. }
\label{S1-Fig}
\end{centering}
\end{figure}

In the first stage of our model the population expands by cell growth and division. A high intracellular growth signal $N_0$ is selected for in the population, favoring fast growing cells (Fig.~\ref{fig_progression}c). This parameter evolves much faster than any other of the 10 mutating parameters (see also Table~\ref{table}) while most of the other parameters do not exhibit such a strong and clear drift by the end of the simulations (Fig.~\ref{fig_progression}d). Indeed, the overall behavior of the model did not change qualitatively in simulations where only $N_0$ or $N_0$ and the chemotactic parameters are allowed to evolve (\nameref{S1-Fig}). Since nutrients in the environment are not limiting due to the assumed prior homeostatis, these cells simply outgrow the slower ones, creating patches of high growth (Fig.~\ref{fig_progression}e). This leads to the expansion of fast growers, and as a result, cells from newer generations appear in clumps of growth hot-spots (Fig.~\ref{fig_progression}f). At this stage expansion can occur at any position in the population since nutrients are available at any location. 

In the second stage of our model the population turns hypoxic. As the number of cells grows, the intracellular pressure increases rapidly in the tightly packed tissue until about t = 5000 MCS (see Fig.~\ref{fig_mutStages}d). At this time oxygen is depleted at areas further away from the source (see Fig.~\ref{fig_progression} middle inset at t=6000 MCS). Fast growing cells (high $N_0$, Fig.~\ref{fig_progression}g) are unable to fuel their increased metabolic need through oxidative respiration and turn hypoxic (Fig.~\ref{fig_progression}h). These cells further increase their glucose uptake (Fig.~\ref{fig_progression}i) due to the HIF1-$\alpha \to$ GLUT signaling pathway in our model (Eq.~\ref{eq:hif}), and start the production of lactate.  

Finally, glucose is gradually depleted at regions far from the sources as a result of the elevated glucose consumption rate of fast growing cells. In the depleted areas cells die out, and with them the cell population is gradually decreased (Fig.~\ref{fig_mutStages}b,d). The only cells remaining are around the vessels, that eventually hijack the source (Fig.~\ref{fig_progression}j, k). These cells continue to compete as in stage 1, since the change in the environment near the vessels is minimal, and keep increasing their internal growth signal from generation to generation (Fig.~\ref{fig_progression}j). Increasing intracellular pressure near vessels (Fig.~\ref{fig_progression}k) exerted by neighboring cells and counteracts growth.

Changes in nutrient levels indicate that our model selects for cells exhibiting the well-known Warburg effect, whereby cells metabolize glucose through glycolysis even in the presence of oxygen (aerobic glycolysis) \cite{VanderHeiden2009}. Cells in our model initially shift to aerobic glycolysis to support their metabolic need escalated through competition, shown by the increasing levels on intracellular hypoxia and ROS (Fig.~\ref{fig_progression}l). This results in an increase in extracellular oxygen (Fig.\ref{fig_progression}a). Despite the availbility of oxygen, cells are unable to revert to a more efficient full respiratory metabolism due to production of ROS which stabilizes HIF1$-\alpha$ and limits the amount of metabolic flux through respiration (Eq.~\ref{eq:hif}), thus keeping it in a state of hypoxia in our model (Fig.~\ref{fig_progression}l). Nevertheless, cells do consume oxygen but it is significantly lower than glucose uptake (Fig.~\ref{fig_progression}m). 

Taken together, these results show that our model exhibits different stages of development similar to previously published studies. Remarkably, this progression emerges in spite of an almost completely unrestricted evolution of a large number of phenotypic parameters. Tumors in this model are initialized at random positions, but due to the explicit representation of localized nutrient sources, we show that they occupy the vicinity of blood vessels at later stages. This is enhanced by the more realistic representation of cells in the CPM where cell shape and compressibility allow cell rearrangements within the packed tissue as opposed to the more rigid CA models exploring progression \cite{Anderson2006, Gerlee2008, Robertson-Tessi2015}. Secondly, we show that our model selects for cells exhibiting the Warburg effect despite the lack of growth advantage of fermenting cells. 

\subsection*{Higher mutation rate speeds up transition between stages}

\begin{figure}
\begin{centering} 
\includegraphics[width=0.9\textwidth]{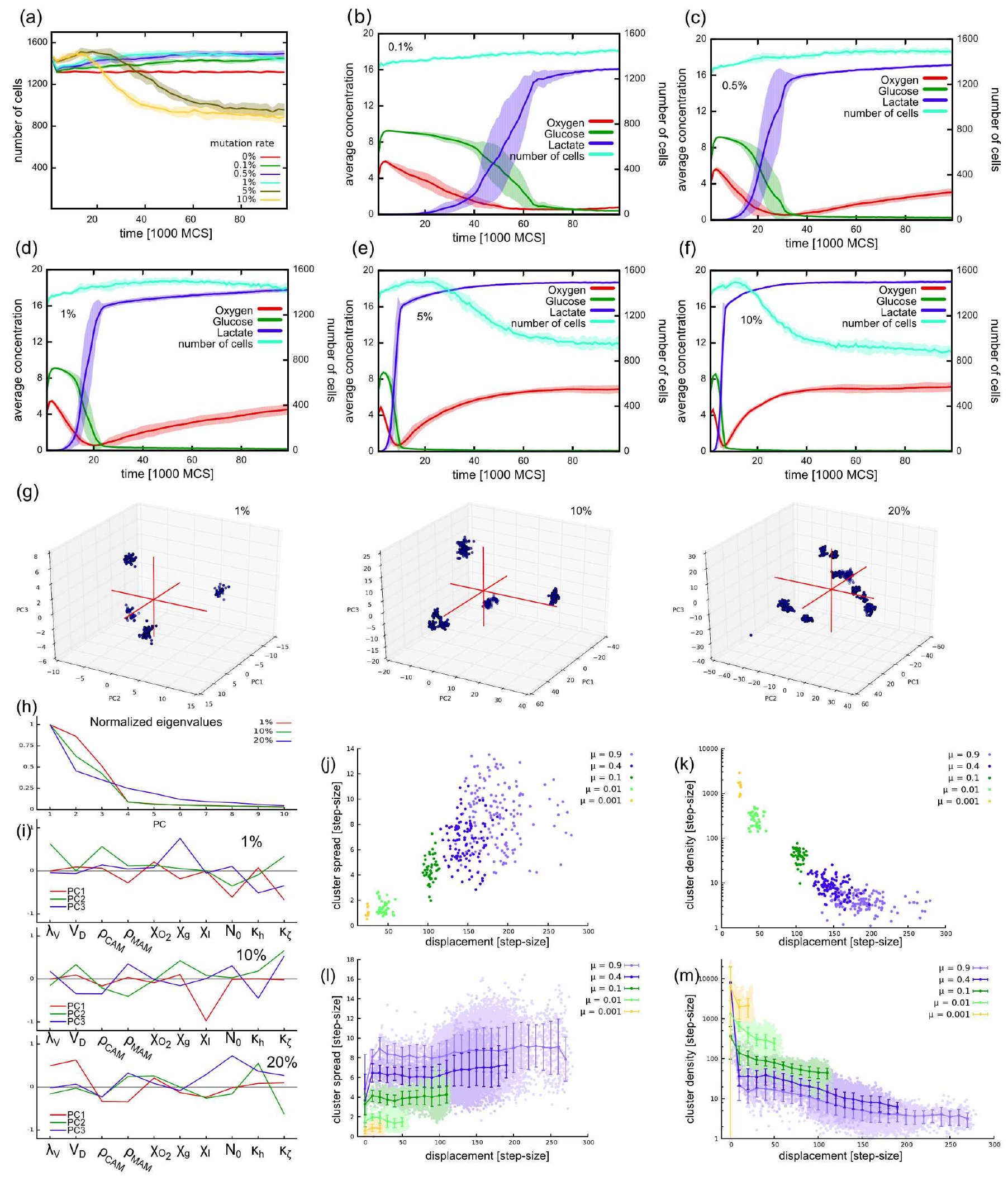}
\caption{\textbf{Effect of mutation rate on population size and stage progression.} (a) Number of cells in the population as a function of time. (b-f) Nutrient levels in simulations with mutation rates 0.1\% (b), 0.5\% (c), 1\% (d), 5\% (e), and 10\% (f) showing a similar behavior in all cases. (g) Distribution of cells along the three main principal components of the populations in phenotype space from example simulations with 1\%, 10\%, and 20\% mutation rates at the end of simulations ($t=10^5$ MCS). Each dot represents a single cell. (h) Relative weight (eigenvalues) of the principal axes of the populations shown in (g). (i) Composition of the first three principal axes in the populations shown in (g). (j-k) Spread and density of clusters identified in phenotype space at the end of simulations using hierarchical clustering, depicted as the function of distance from the origin. Each dot corresponds to one cluster from a total of 10 independent simulations. (l-m) Spread and density of clusters throughout the evolution of the model. Data from 10 independent simulation repeats from each condition, except (g-i) which shows data from one simulation from each condition.} 
\label{fig_progVsMutRate} 
\end{centering} 
\end{figure}

\begin{figure}
\begin{centering}
\includegraphics[width=1.00\textwidth]{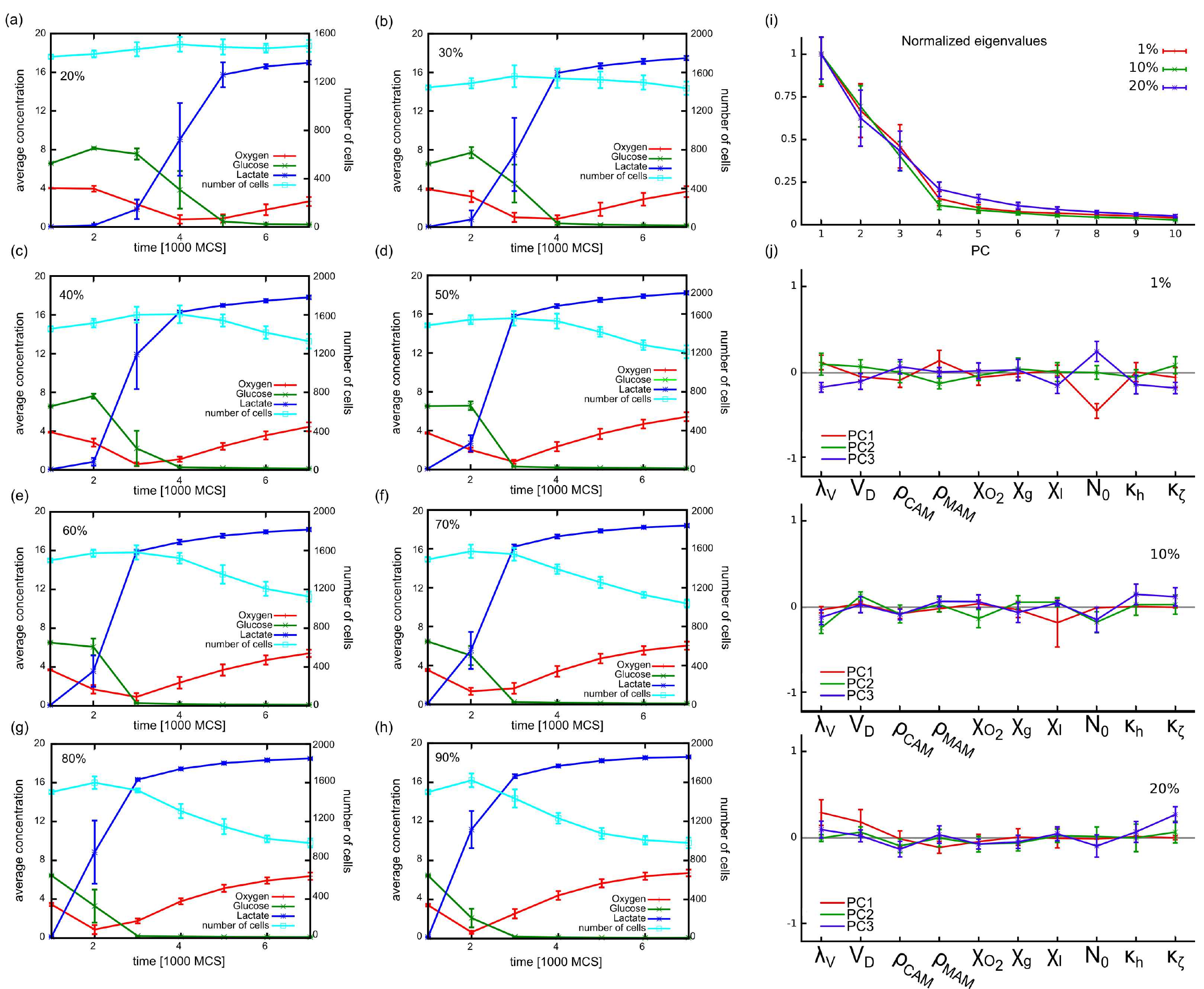}
\caption*{{\bf S2 Fig. Effect of mutation rate on population size and stage progression.} Average number of cells and nutrient concentrations in simulations with mutation rates of 20\% (a), 30\% (b), 40\% (c), 50\% (d), 60\% (e), 70\% (f), 80\% (g), and 90\% (h). Progress of populations through stages is increasing with increasing mutation rates. Population averages from 10 independentent simulation runs with standard deviation accross simulations. (i) Relative weight (eigenvalues) of the principal axes of populations in phenotype space averaged from 10 independent simulation runs for 1\%, 10\%, and 20\% mutation rates each. (j) Averaged composition of the first three principal axes in the populations shown in (i).}
\label{S2-Fig}
\end{centering}
\end{figure}

To test if the stages of tumor progression depend on the phenotypic mutation rates, we simulated the model for a series of mutation rates. Whereas the non-mutating population keeps a constant size, all mutating populations exhibit an initial increase in cell numbers (Fig.~\ref{fig_progVsMutRate}a). In highly mutating populations (5\% and 10\%) this increase is followed by a drop in cell numbers. This drop is observed later in populations with 5\% mutation rate, and population decrease is just starting at the end of the simulations in populations with 1\% mutation rate. Note that the repetitions reproduce the behavior fairly well, suggesting the robustness of the system. Therefore we suggest that similar stages occur at lower mutation rates, and the time needed for reaching each stage depends on the mutation rate. This is supported by the changes in nutrient levels in the simulations, which react faster to change than the total number of cells (Fig.~\ref{fig_progVsMutRate}b-f). In healthy, non-mutating populations, nutrient levels and population size is stabilized (Fig.~\ref{fig_stableStages}b). In mutating populations, cells evolve a higher metabolic demand in parallel with increased proliferation. This results in an increase in population size and decrease in oxygen levels, followed by a decrease in glucose levels. As oxygen becomes sparse, cells turn hypoxic and switch from oxydative phosphorylation to aerobic glycolysis, resulting in an increase in lactate levels. We have found the same behavior in populations with different mutational probabilities ranging from 0.1\% up to 10\% (Fig.~\ref{fig_progVsMutRate}b-f), or higher (\nameref{S2-Fig}~a-h). Again, a lower mutational probability only delayed the changes in the nutrient levels. We did not find a qualitative difference in the progression in our simulations at different mutation rates, showing that the emergent order of stages is robust in this system.

To explore the structure of the population at different mutation rates in the system, we analysed the distribution of cells in the space of normalized mutating parameters. After subtraction of the initial parameter values and normalizing with mutational step-size, the ten-dimensional parameter space of the population was reduced to the three most prominently changing axes within each population using principal component analysis (see Methods). Fig.~\ref{fig_progVsMutRate}g shows one example population at the final time point of the simulations for mutation rates 1\%, 10\%, and 20\% with each dot representing a cell. At low mutation rate (1\%) the population splits up into well-defined clones which are more spread and less well-defined at higher mutation rates. The first three principal axes contain most of the information about the shape of the population, as can be seen by the normalized weights (eigenvalues) of these components (Fig.~\ref{fig_progVsMutRate}h). The composition of principal axes at the final time point of the simulations shown on Fig.~\ref{fig_progVsMutRate}g differ (Fig.~\ref{fig_progVsMutRate}i) with the growth signal $N_0$, cell rigidity $\lambda_V$, and glucose chemotaxis $\chi_g$ playing an important role at $\mu=$1\%. At 10\% mutation rate the lactate chemotaxis parameter $\chi_l$ plays a distinct role in segregating the population phenotypes, while at $\mu=20\%$ the segregation is less obvious (Fig.~\ref{fig_progVsMutRate}g) and is driven mainly by parameters $\lambda_V$, doubling volume $V_D$, and adhesions ($\rho_{\text{CAM}}, \rho_{\text{MAM}}$). While the composition of the main axes varies across different simulation repeats with the same mutation rate, the populations are nevertheless well characterized by the first three components in all cases (\nameref{S2-Fig}~i, j). 

To better understand population structure in the simulations, we categorized the cells at each time point in the 10-D phenotype space using hierarchical clustering (Methods). We measured the displacement of each cluster as the Euclidean distance between the point of origin and its center of mass and its spread as the mean distance of points of the cluster from the cluster's center of mass. As expected, we observed that populations with higher mutation rates reach further from the origin by the end of simulations; these clusters are more spread and less dense than clusters in populations with lower mutation rates (Fig.~\ref{fig_progVsMutRate}j, k). Considering the whole time course of the simulation, the cluster analysis reveals that as the population explores the phenotype space, clusters from the highly mutating populations first tend to spread out and dilute more than the clusters in populations from lower mutation rates (Fig.~\ref{fig_progVsMutRate}l, m). These results show that the population starts to dilute much faster in phenotype space at high mutation rates, but without affecting the progression of stages apparent from the nutrient levels.

\subsection*{Faster growing cell population is less robust in the face of external nutrient fluctuations}

Next, we tested how feedback from a larger spatial organizational level, through the nutrient supply in our case, would affect populations of different mutation rates. In a healthy tissue, nutrient supply is relatively constant. The main source of fluctuations are the slow daily change according to the circadian rhythm, and the relatively fast blood pulse. In cancerous tissues the vasculature is remodeled through tumor vasculogenesis, resulting in tortuous and leaky vessels \cite{Carmeliet2011,Weis2011}. As these vessels are less reliable, here we assume that they dysfunction from time to time, for example by becoming temporarily blocked.

To model vessel tortuosity, we introduced a blocking probability for the vessels. An open blood vessel will be blocked with a probability $\mathcal{P}$ at every time step, and a blocked vessel will be opened with the same probability. A high blocking probability ($\mathcal{P}=0.5$) corresponds to a healthy situation, where the resulting fast switching of the vessel is smoothened out by nutrient diffusion. A lower blocking probability introduces longer periods of nutrient deprivation but also longer periods of nutrient supply. On average these systems receive the same amount of nutrients, but in different dosage. 

\begin{figure}
\begin{centering} 
\includegraphics[width=0.90\textwidth]{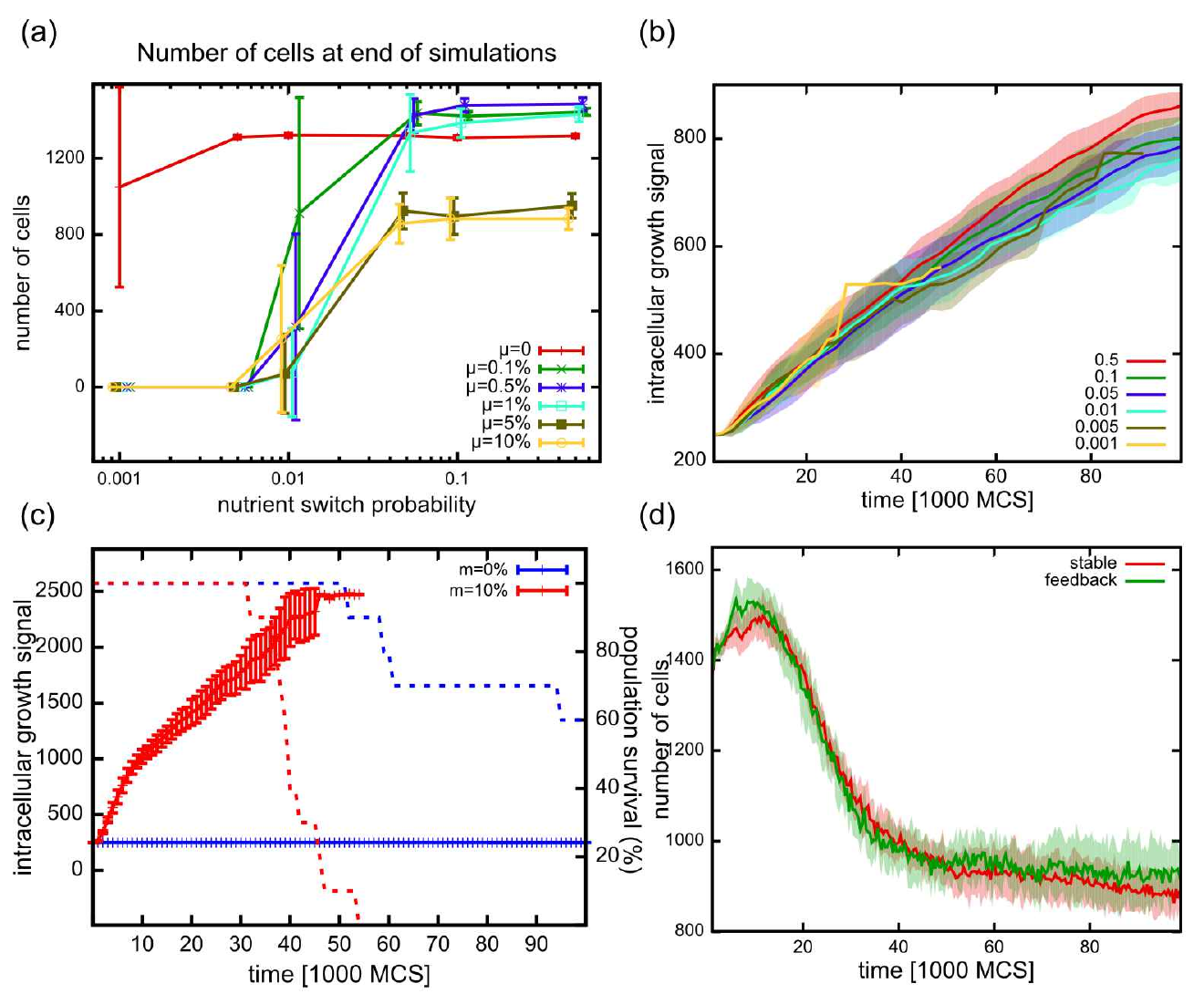}
\caption{\textbf{Effect of fluctuating nutrient supplies on the tissue.} (a) Number of cells at the end of simulations (t=100,000 MCS), showing for simulations with different vessel tortuosities ($\mathcal{P}$) and different cell mutation rates ($\mu$). In simulations with healthy vessels (high switching probability $\mathcal{P}$) a mutating cell population produces more cells than non-mutating ones. However, in simulations with erratic nutrient supply (with lower blocking probability $\mathcal{P}$), the mutating populations die out more frequently than the non-mutating ones. At extreme erratic switching ($\mathcal{P}=0.001$) 10 out of 10 mutating populations are extinct and even healthy ones start to die out (2 out of 10 simulations). Error bars indicate standard deviation of cell numbers from 10 repetitions. Values belonging to different simulations with the same mutation rate are connected and are slightly shifted on the x-axis for better visibility. (b) Average intracellular growth signal evolution in the population are not changed in simulations with different vessel blocking probabilities (mutation rate $\mu=0.1\%$). This shows that the vessel blocking probability does not directly affect selection and progression speed. (c) The average intracellular growth signal in simulations with decreasing vessel blocking probability ($\mathcal{P}=\mathcal{P}(t)$, solid lines) in mutating (red, $\mu=10\%$) and not mutating (blue, $\mu=0\%$) populations. Cell-cell competition in mutating populations initially drives the growth parameters to a high value just as in the simulations with stable nutrient sources. When the blocking probability reaches the magnitude of $\mathcal{P}(t)=0.01$ and below, populations die out (dashed lines: percentage of simulations with living cells). (d) Population sizes in simulations with stable nutrient sources ($\mathcal{P}=0.5$) and in simulations where the blocking probability is controlled by the cell density ($\mathcal{P}=\mathcal{P}(\rho)$), creating a feedback. In both cases the populations survive. The tissue coverage is approximately halved, making the vessel blocking probability approximately $\mathcal{P}=0.27$ in the feedback simulations. } 
\label{fig_fluctSource} 
\end{centering} 
\end{figure}

Healthy, non-mutating populations in our simulations survived over a wide range of nutrient fluctuations. The average number of cells from 10 simulation repeats showed that these populations keep roughly the same size at different blocking probabilities, as shown on Fig.~\ref{fig_fluctSource}a. At the extreme blocking probability $\mathcal{P}=0.001$, only 2 out of 10 populations died out. 

In comparison, mutating cell populations are unable to tolerate blocking probabilities lower than $\mathcal{P}=0.01$, irrespective of their mutation rate (Fig.~\ref{fig_fluctSource}a). Note that at high $\mathcal{P}$, populations with low mutation rates have an increased population size at the end of the simulations ($t=10^5$ MCS), compared to the healthy population. This increase results from the initial stage of progression, as these populations only reach the first stage (expansion) by the end of the simulations. However, this advantage disappears as $\mathcal{P}$ decreases. 

The observed reduction in population size due to decreasing $\mathcal{P}$ could work in two ways: either by killing cells through starvation, or by speeding up the progression of stages. In the previous section we showed that a higher mutation rate speeds up the progression of the population and thus results in a reduced population size (Fig.~\ref{fig_progVsMutRate}). This might eventually lead to extinction. If the populations under fluctuating nutrient supply go through the same stages of progression as the ones in the stable environment, the environmental indicators used in Fig.~\ref{fig_progVsMutRate} (levels of glucose, oxygen, lactate) are unsuitable, as these might change in simulations with different $\mathcal{P}$. Instead, we focus on the intracellular evolution of traits, that are not altered directly in these experiments. The first trait to be selected for is the intracellular growth signal of the cells ($N_0(i,t)$) that exhibits a run-away dynamics (Fig.~\ref{fig_progression}c). If the blocking probability accelerates the progression through the stages, it should increase the selection pressure on the intracellular growth signal as well. Contrary to this expectation, we found that the trend in the average value of the intracellular growth signal remains approximately the same in simulations across different $\mathcal{P}$ values, and even slightly decreases at lower $\mathcal{P}$ (Fig.~\ref{fig_fluctSource}b). Similarly, other cellular measures (such as hypoxia, ROS, or pressure), or cellular parameters (such as the chemotaxis parameters) show the same behaviors irrespective of $\mathcal{P}$ (\nameref{S3-Fig}). Therefore, longer nutrient fluctuations do not accelerate the evolution of the population, and the reduction in population size is not a result of the acceleration of the same evolutionary dynamics. Instead we conclude that as cells deplete the nutrients in the environment due to their increased consumption, the chance for survival in systems with longer fluctuations is reduced. 

Next we altered $\mathcal{P}$ during the time of simulation runs. We tested how the population reacts if the blood vessels become increasingly tortuous, starting from a healthy state (fast switching) progressing to a tortuous vasculature (slow switching). Blocking probability in these simulations is decreased gradually in the simulations, following a geometric progression $\mathcal{P}(t+1)=r\mathcal{P}(t)$ with an initial value of $\mathcal{P}(t=0)=0.5$ and ratio of $r=0.999876$. Once the progression reaches $\mathcal{P}(t_f)=0.001$ at $t_f\approx50,100$ MCS the blocking probability is not decreased further ($\mathcal{P}(t>t_f)=\mathcal{P}(t_f)$). Similar to the stable system, the mutating populations (mutation rate $\mu=0.1$) are initially driven into the high consumer state by cell-cell competition (Fig.~\ref{fig_fluctSource}c solid red line showing growth signal parameter $N_0$). When the fluctuation probability reaches the order of $\mathcal{P}(t)=0.01$ ($t\approx 31,500$ MCS), the populations start to die out (Fig.~\ref{fig_fluctSource}c red dashed line showing number of surviving populations), similar to the case of static low blocking probabilities. Note however, that non-mutating populaions (Fig.~\ref{fig_fluctSource}c blue) are able to survive increasing fluctuations in the nutrient supply. This shows that a changing nutrient supply does not necessarily influence the direct competition among cells. 

Inconsistency of nutrients may emerge from the disfunctional tissues of the emergent tumor occluding vasculature.  To represent this feedback, we examined how the population behaves when the consistency of nutrient supply is related to the amount of cellular coverage in the tissue. We created a feedback between the density of the tissue (measured as tissue surface coverage $\rho$, with $0\leq\rho\leq 1$) and the fluctuating source. For a fully populated tissue we kept the nutrient switching probability high, $\mathcal{P}=0.5$, providing a smooth nutrient supply, and decreased it linearly with the cell density to model the variability in nutrient supply. Thus: $\mathcal{P}(\rho) = 0.499\rho + 0.001$. In these simulations the mutating population persists at approximately the same level as in the healthy case (Fig.~\ref{fig_fluctSource}d). The emergent tissue coverage yields an approximate fluctuation probability of $\mathcal{P}(\rho)\approx0.27$, sufficiently high to support mutating populations. While the relationship between nutrient supply stability and living cell density is experimentally unclear, our results suggest that a linear relationship between stability and density is insufficient to drive the agressive cells to extincition, similar to what is expected of an expanding pathological tumor.

The effect of low nutrient levels with two consecutive surges of high oxygen levels has been shown to affect selection in a model of evolutionary tumor growth \cite{Anderson2006}. Here we used localized nutrient sources and stochastic supply, rather than two deterministic and uniform pulses. We show that in our system, inconsistent nutrient supply does not increase the speed of cellular evolution, however, it does reduce the viability of the unstable cell populations. We found similar results when inconsistency is considered progressively increasing or proporitional to the tissue coverage. Our results suggest that, while consistent nutrient supply promotes cell-level selection of fast growing cells, inconsistent nutrient supplies that put a larger demand on the tissue exert selection at the tissue scale and provide higher chances of survival for populations with cellular quiescence.

\subsection*{Clonal expansion and heterogeneity as a consequence of differential growth}

\begin{figure}
\begin{centering} 
\includegraphics[width=0.90\textwidth]{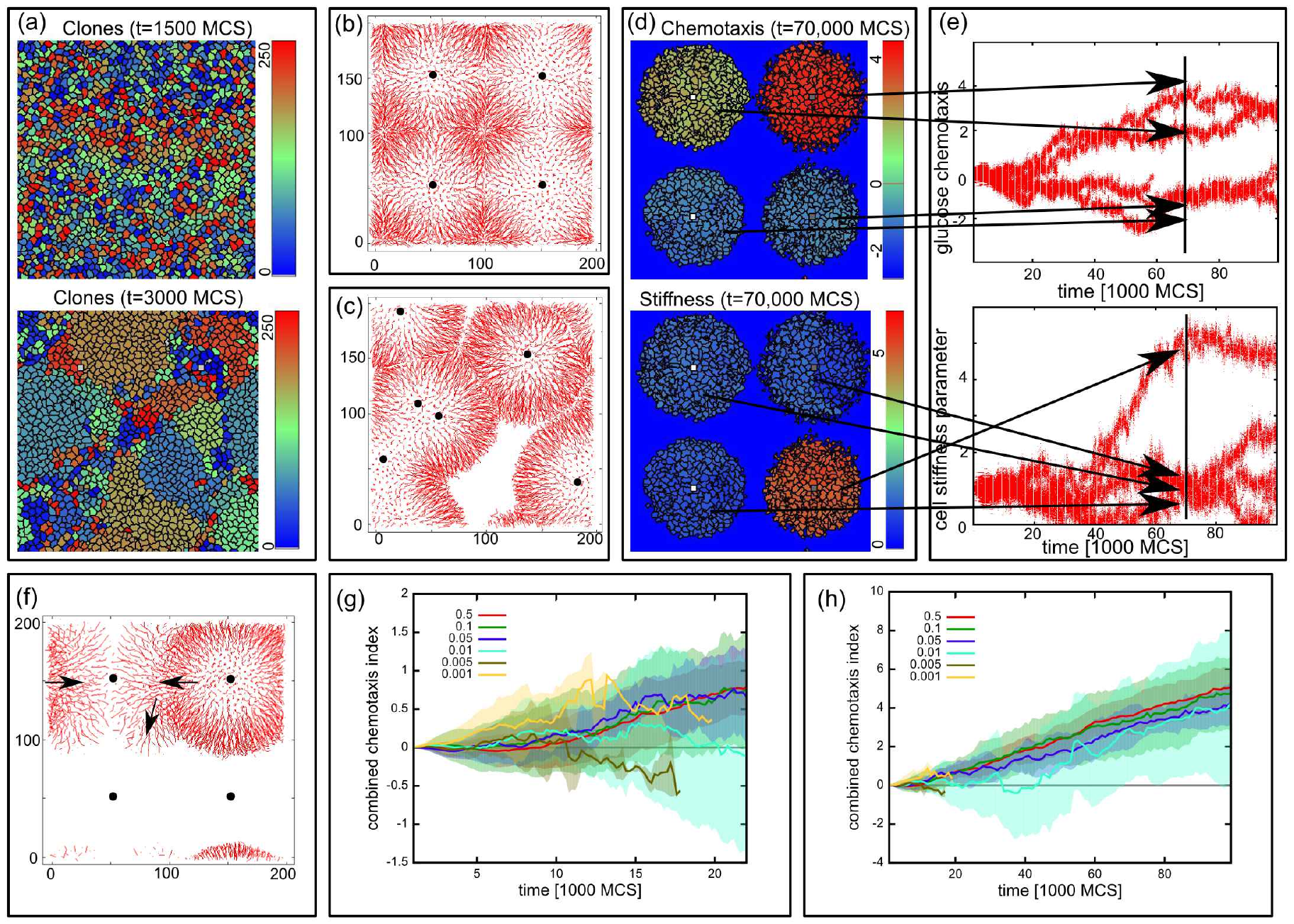}
\caption{\textbf{Clonal segregation.} 
(a) Clonal expansion. Model configurations of the same population at t=1500 MCS and t=6000 MCS. Each cell is depicted with one of 256 different colors at time t=1500 MCS, and the daughters inherit this color from that time point onwards. The population 4500 MCS later consists of the descendants of about a dozen of the initial cells. (b) Cell trajectories showing that cells flow outwards from the nutrient sources (indicated with black dots). (c) Trajectories from a setup with randomly placed vessels, showing that the outflow is determined by the placement of vessels. (d) The population segregates into independently evolving parts around the blood vessels. Color code shows the chemotaxis parameter for glucose ($\chi_g(i,t)$) and the stiffness parameter $\lambda_v(i,t)$ at t=70 000 MCS. Cells around the same source are homogeneous, but are different from cells surrounding other sources. (e) Chemotaxis parameter ($\chi_g(i,t)$) and stiffness parameter ($\lambda_v(i,t)$) values in the populations shown in (d) as the function of time. The four segregated parts clearly separate in their parameter values, and evolve almost independently. Vertical line denotes the time t=70 000 MCS corresponding to the time of the configurations on (d). (e) Trajectories from a late population in the fluctuating system, showing the trajectories moving from one source to another. Movement direction is indicated with arrows. (g, h): Combined chemotaxis parameters ($\langle\chi'(i,t)\rangle_i=\langle\chi_g(i,t) + \chi_o(i,t) - \chi_l(i,t)\rangle_i$) as a function of time in simulations with different vessel tortuosities at the first 20 000 MCS (g) and for the whole simulation time (h). The parameter is increasing in all simulation conditions, showing that nutrient detection is important for cell survival. In simulations with lower blocking probabilities the random selection introduced by the blocking and opening of the vessels results in a higher noise in the combined chemotaxis parameters average. Values averaged from 10 independent simulation runs for each condition. } 
\label{fig_segregation} 
\end{centering} 
\end{figure}

In a simulated healthy tissue devoid of mutants, cell growth is independent of spatial localization, growth is limited intrinsically by a constant growth signal balancing spatial confinement. After this intrinsic limitation is lifted through uncontrolled mutations and competition, growth becomes clustered in growth hot-spots (Fig.~\ref{fig_progression}e). These spots are populated by overly proliferative cells, resulting in clonal expansion: Fig.~\ref{fig_segregation}a shows configurations at two time points in the same simulation color-coded for descendants. At a later stage, when cells deplete nutrients, proximity to the sources dictates the growth rate (Fig.~\ref{fig_progression}j). In the resulting environment cells closer to the source grow faster and divide, while cells further away starve and die. This differential growth gives rise to a directed cell movement from the vicinity of the sources to the depleted areas, apparent from the short cell trajectories shown in Fig.~\ref{fig_segregation}b. To demonstrate that the segregation patterns were indeed linked to the nutrient sources and are not the result of finite system size, we performed simulations with randomly scattered blood vessels instead of the regular distribution, similar to previous studies \cite{Scott2016,Powathil2012}. The example trajectory plot on Fig.~\ref{fig_segregation}c shows that the outflow of cells is correlated with the nutrient positions. 

The population becomes segregated, as the probability of a cell invading a neighboring region is diminished. As a result of this differential growth cells near the source take over the vicinity of the vessel and spread their phenotype in this region (Fig.~\ref{fig_segregation}d). While cells within each segregated part of the population have highly similar phenotype, the different parts evolve independently of one another. This is shown by the distribution of phenotypic traits over time in each population (Fig.~\ref{fig_segregation}e). This gives rise to independently evolving quasi-species-like populations within the tissue, and might be analogous to the observed heterogeneity in tumors.

In simulations with low but constant vessel blocking probabilities cells around a blocked source are able to leave their region and migrate to another active source, as exemplified by the trajectories in Fig.~\ref{fig_segregation}e. The ability to detect and to move to a neighboring active source is crucial for the cells for survival, and therefore is expected to be selected for. In our model, cells can achieve this by evolving chemotaxis towards glucose or oxygen, or chemotaxis away from lactate. Indeed, chemotaxis parameters are the second most affected parameters throughout the evolution of the population (Fig.\ref{fig_progression}c, d). Interestingly, chemorepulsion by lactate is strongly selected for although lactate in our model does not have any direct metabolic effect on the cells. This acquired property helps to orient the cells towards the nutrient sources better than oxygen, which becomes ubiquitous with more shallow gradients, or glucose, which does not diffuse far from the nutrient sources due to elevated uptake. The sum effect of these motility parameters can be expressed with a combined chemotaxis parameter defined as: $\chi'(i,t)=\chi_g(i,t) + \chi_{O_2}(i,t) - \chi_l(i,t)$. Fig.~\ref{fig_segregation}g and h show the evolution of this combined chemotaxis parameter in populations from different vessel blocking probabilities and mutation rate of 10\%. Indeed in all of the conditions the combined chemotaxis is selected for. In simulations with lower blocking probabilities the combined chemotaxis has increasingly higher fluctuations due to the random selection introduced by the blocking and opening of the vessels, often leading to extinction as well.

\section*{Discussion}
In this study we introduced a computational model to explore the evolution of tissue cells under specific conditions. Selection in the model acts through physical constraints: limited growth space, and limited nutrient diffusion. Selected cellular phenotype traits, such as cell rigidity or growth signal, are allowed to change (``mutate") upon cell division in any direction with a probability within a given range. 

\subsection*{Progression of phenotypes}

We show that a directional evolution emerges from random movement in phenotype space as the result of cell competition, driving cells from a healthy (homeostatic) state to a more aggressively expansive phenotype. This is consistent with previous findings of Anderson and colleagues \cite{Anderson2006}, who showed that a similar drift towards aggressive phenotypes emerges if cells are allowed to mutate randomly into one of a 100 predefined discreet phenotypes. In contrast to these abrupt changes in behavior, our model only allows small changes in mutation. This choice lends more persistence to the clones in the population, since more mutation events are required to diverge from an existing clone. Faster growing cells are selected in our model, which then go on to colonize the population by means of clonal expansion. 

Progression of phenotypes has been observed previously in other models of tumor evolution, where the authors also considered the toxic effect of acidification due to glycolysis \cite{Gerlee2008,Robertson-Tessi2015}. In \cite{Gerlee2008}, cells first develop the ability to survive in hypoxic environments by lowering their apoptitic response threshold to low oxygen levels. This is followed by a metabolic switch to glycolysis and finally the emergence of the acid-resistant phenotype. In contrast, acid-resistance was proposed to emerge first followed by evolution of glycolysis in another model where nutrient sources were represented as point sources \cite{Robertson-Tessi2015}. The distinct stages exhibited in our model (clonal expansion, oxygen depletion, and starvation) are accompanied by a sudden increase in glycolytic activity which is then moderated onto a stable medium level (Fig.~\ref{fig_progression}l). In our model the population turns hypoxic (Fig.~\ref{fig_progression}l) before developing a distinct chemo-repulsive response (Fig.~\ref{fig_progression}c). We found the progression robust against changing the mutation rate, however, at higher mutation rates the population was able to explore the phenotype space faster and in more dispersed clusters (Fig.~\ref{fig_progVsMutRate}). 

Although lactate in our model does not affect cells in our model, cells gradually develop a chemorepulsion away from lactate (a negative $\chi_l$ on Fig.~\ref{fig_progression}c, d). Lactate accumulates at regions where high consumers deplete glucose, therefore cells use lactate chemorepulsion as a compass to navigate away from the high-consumer niche towards a more supportive environment. This novel feature in our model highlights how phenotypes that show no apparent advantage at the cell level could be selected for based on the altered micro-environmental conditions. 

\subsection*{Minor differences in proliferation lead to clonal expansion}

Local expansion in tumor growth models has recently been described by Waclaw and colleagues \cite{Waclaw2015}. Their study focused on heterogeneity in passanger mutations while the probabilistically occurring driver mutations set the proliferation advantage. Cell motility there acts to blur intratumor heterogeneity by allowing proliferating cells to invade `cell-free' areas where further proliferation is not inhibited by other cancer cells. While this model is able to reproduce the clonal expansion features of tumors, it neglects the potential inhibitory effect of spatial constraint produced by the surrounding stromal tissue. Proliferation diversity in tumors was shown to be essential to explain experimentally observed tumor morphology by an earlier study using the cancer stem cell hypothesis \cite{Enderling2009} which interprets the tumor as a conglomerate of self-metastases. The emergent clonal expansion patterns in our model (Fig.~\ref{fig_segregation}a) are reminiscent of these self-metastases and therefore suggest that slight differences of proliferation rates within the population are sufficient to generate these patterns, as opposed to the sharp distinction between cancer stem cells and non-stem cells. 

\subsection*{Spatial confinement and localized nutrient sources}

Previous models of tumor evolution typically neglect the confining effect of the surrounding healthy tissue although spatial confinement can play an important role when studying the effects of treatment recovery in a tumor with a cancer stem cell population \cite{Sottoriva2010}. When a large portion of cells is killed by therapy, the internal cells have access to growth space and are able to regrow the tumor. In other words: space limitation keeps the growth inhibited. To mimic this situation here we restrict the growing tumor to a confined space, making them a model of an in vivo system. We note that simulations in our system do not develop a necrotic core as in other models, as dead cells simply shrink and disappear from the simulations without inducing any signaling response from neighboring cells or without increasing spatial confinement. Therefore our system focuses on live tumor cell evolution where necrotic cells may be considered as extruded from the simulated monolayer region into the underlying necrotic core, removed by the immune response and/or drained by the lymphatics. 

The source of nutrients in previous models is typically considered uniform, or is provided in the environment in a dispersed fashion. The source of nutrients are the blood vessels in our model, similar to the more recent studies of Shirinifard and colleagues \cite{Shirinifard2009}, Powathil and coworkers \cite{Powathil2012, Powathil2014, Powathil2016}, or using a hybrid cellular automaton model \cite{Patel2001}. Two recent studies showed that arrangement of vessels and their 2D representation affects tissue oxygenation and may alter the outcome of a simulated radiotherapy \cite{Grogan2016, Scott2016}. Nutrient diffusion and utilization create a spatial gradient in the supplies which is translated into a differential growth pattern in our system. Differential growth creates a collective flow of cells outwards from the sources (Fig.~\ref{fig_segregation}b, c, f). This outward flow is counter-acted by the emergent chemotactic tendency of cells to move closer to the source (Fig.~\ref{fig_progression}c, d). Chemotaxis in a changing environment provides additional advantage in locating active nutrient sources (Fig.~\ref{fig_segregation}f). 

Blood vessels in our model are immobilized entities, vascular remodeling is not included. Inconsistency in nutrient supply is implemented as stochastic switching of blood vessel activity, but without feedback from the pressure or hypoxia in the tissue as in the phase field model of Yan and colleagues \cite{Yan2016}. Simulations where the blocking probability $\mathcal{P}$ is a function of time (Fig.~\ref{fig_fluctSource}c) or tissue coverage (Fig.~\ref{fig_fluctSource}d) showed that the progression of stages is not altered in our system. A higher number of connected loops have been shown to emerge in 3D models of tumor angiogenesis than in 2D, predicting that complete blockage of circulation is very rare \cite{Stephanou2005, McDougall2006}. Here we studied local vessel blockage which would still have a local effect on our system, even if at a reduced frequency. 

\subsection*{Inconsistent nutrient supply}

\begin{figure}
\begin{centering}
\includegraphics[width=1.00\textwidth]{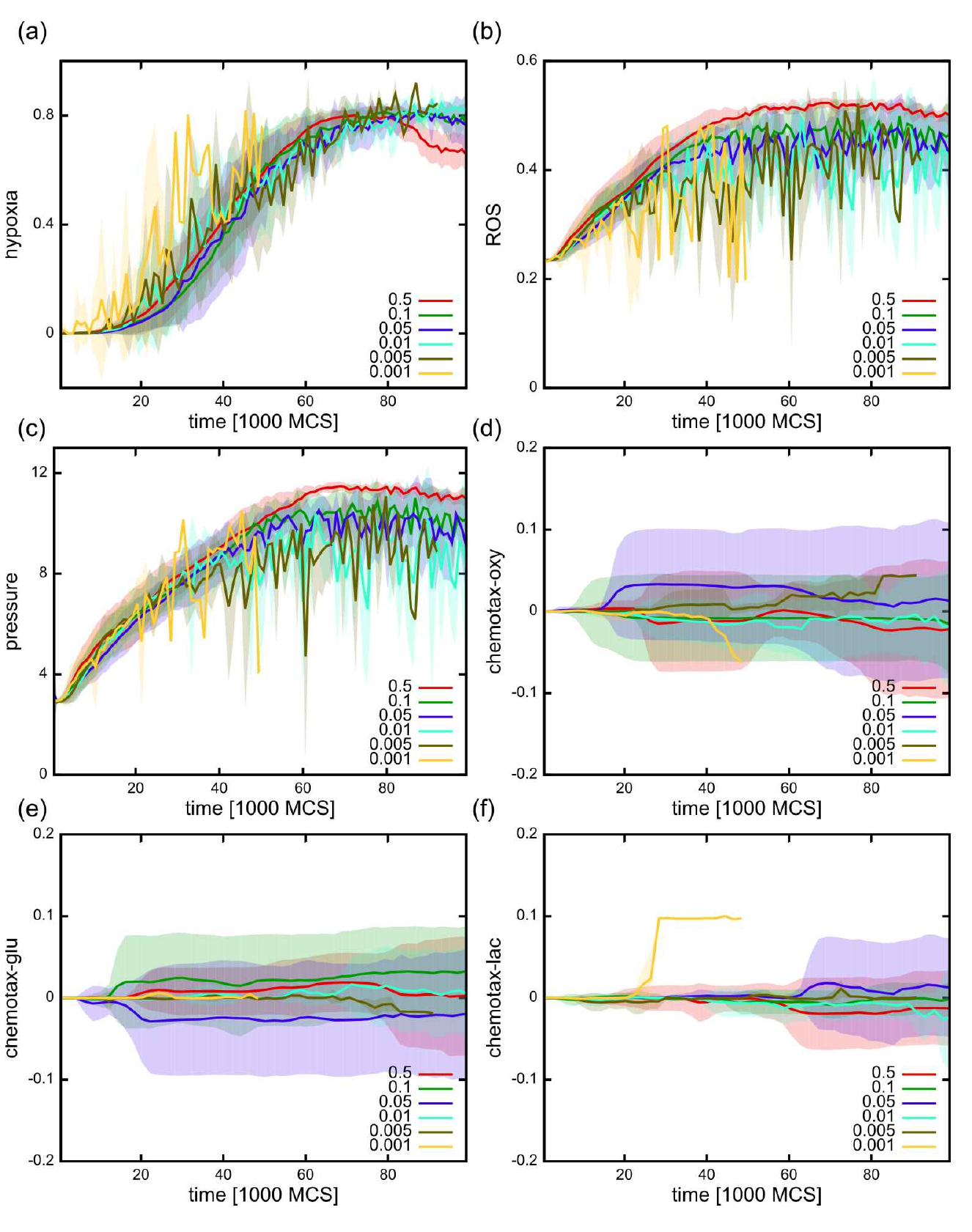}
\caption*{{\bf S3 Fig. Effect on fluctuating nutrient supply on populations with 10\% mutation rate $\mu=0.1\%$.} Hypoxia (a), ROS (b), cellular pressure (c), and chemotaxis towards oxygen (d) glucose (e) and lactate (f), remain largely unaffected by different vessel blocking probabilities. Note that the variation increases with lower blocking probability $\mathcal{P}$ due to fewer surviving cell populations. Population averages from 10 independentent simulation runs with standard deviation accross simulations. Blocking probability values $\mathcal{P}$ show on graphs.}
\label{S3-Fig}
\end{centering}
\end{figure}

Previously Anderson and colleagues showed that surges of oxygen induce diversification into a population of simulated cells evolving under low oxygen levels \cite{Anderson2006}. A repeated, second surge was shown to induce phenotypic segregation in these simulations \cite{Anderson2006}. Here we studied how inconsistency of the nutrient supply affects the population using stochastic (rather than deterministic) switching. In contrast to Anderson's study, cells do not receive a constant low oxygen supply in our model and are therefore prone to extinction. We found that populations with increasing mutation rate are increasingly sensitive to nutrient fluctuations (Fig.~\ref{fig_fluctSource}a), although the evolutionary trend at the cell level remained largely unaffected (Figs.~\ref{fig_fluctSource}b, \nameref{S3-Fig}). In our model the aggressive phenotype analogous to cancer is a natural consequence of selection on the cell-level and random mutation. When selection pressure is applied on the tissue level (in the form of fluctuating nutrient supplies) the overall fitness (or survival) of the cancerous population proves to be lower than that of the healthy population, potentially due to depletion of ambient nutrient resources as a form of competition. Cancerous cells create an insecure environment that is more sensitive to stress coming from outside the cell population.

This theoretical exploratory study opens questions in how to best approach renormalization of the tumor vasculature. In the model, healthy cells are able to withstand the irregularities in nutrient supply better than cancerous cells. Based on this observation, it is tempting to speculate that irregularities of the tortuous tumor vasculature might serve a similar role in real tumor development. It is important to bear in mind that these results are based on a rudimentary model of tumor development. In addition to the simplifications discussed above, cell-cycle regualtion is overly simplified in our model as opposed to the study of Powathil and colleagues \cite{Powathil2012} where it is in the focus of the study; cells in our model lack an explicit way to store surplus energy as opposed to the cumulative health-factor of Swat and co-workers \cite{Swat2015} representing the cells' tolerance against starvation. Due to its 2D nature, our model is unable to account for the out of plane transport of nutrients or cells. However, using our quasi-2D model allows the study of several processes taking place in epithelial tissues, where most of the tumors arise. Importantly, the implemented cellular metabolism is overly simplified to enable the exploration of the system. After the foundation of this model framework is set out, it could be expanded with more detailed intracellular metabolic networks, for example, using spatial dynamic flux-balance analysis \cite{vanHoek2017}. These multiscale models will lead to a tighter integration of computational and experimental work similar to the ``symbiotic'' approach described in a recent angiogenic sprouting study of Cruys and colleagues \cite{Cruys2016}. 

Further future work includes the more thorough exploration of different mutation rates for different traits. One of the ultimate goals of computational systems biology is constructing a virtual tissue in order to predict efficient treatment of various diseases. This can be achieved by adding homeostatic mechanisms to the tissues such as contact inhibition of growth, or introducing an internal energy storage as in the study of Swat and colleagues \cite{Swat2015}. Inclusion of a dynamic angiogenesis model (e.g. ~\cite{McDougall2006, Pries:2009fx, RMHMerks:2008fv, Shirinifard2009, Szabo:2008em}) might involve exploring the roles of different blood vessel placements, or implementing angiogenesis models already available in the same platform. Finally, the system enables the measurement of fitness at different spatial levels and in different (micro- and macro-) environments which could serve as a basis for exploring evolutionary trade-offs in using computational simulations. 

\subsection*{Conclusion}
In this study we introduce an unbiased evolutionary approach to studying the evolution of interacting tissue cells. The model includes localized source of nutrients (oxygen, glucose) and sink for intermediate metabolites (represented by lactate in our model), and a simplified cellular metabolism including glycolysis and respiration. Cells in the model are spatially confined and no explicit distinction is made between stromal or tumor cells. The model exhibits distinct stages of development with an emergent evolutionary drift towards rapid growth, high glucose consumption, and hypoxia. This is accompanied by a Warburg-effect, whereby cells become unable to return to respiratory metabolism even at high oxygen levels. The simulated tumor exhibits clonal expansion and eventually gives rise to similar phenotypes around each nutrient source which then evolve independently later on. Finally, we found that the emergent rapid growing population is highly sensitive to intermittent nutrient supply, such as caused by leaky tortous blood vessels. 

\section*{Methods}

\subsection*{Details of computational model}
\subsubsection*{Cell volume, adhesion and chemotaxis}

The CPM was introduced in the early 1990's \cite{Graner1992, Glazier1993} and has evolved into a widely used computational model, applied to various phenomena in developmental biology, cell biology, and tumor biology. Cells in the model are represented as domains on a regular square lattice $\Lambda \subset \mathbb{Z}^2$ with an integer spin, $\sigma(\vec{x})\in\mathbb{Z}^{+,0}$, on each lattice site $\vec{x} \in \Lambda$ that uniquely identifies the biological cell occupying that site, with $\sigma(\vec{x})=0$ identifying the extracellular region (medium). Cell movement is described as a series of elementary steps, in which an attempt is made to copy $\sigma(\vec{x})$ of a randomly selected site $\vec{x}$ into an adjacent site $\vec{x}' $. The attempt is accepted with probability 1 if it decreases the value of a globally defined Hamiltonian function, $H$, or with Boltzmann probability if it would increase the value of $H$:

\begin{equation}
 p(\sigma(\vec{x})\to \vec{x}')=\begin{cases} 1 & \mbox{ , if } \Delta H(\sigma(\vec{x})\to \vec{x}') < 0 \\
   \exp{\left[\cfrac{\Delta H (\sigma(\vec{x})\to\vec{x}')}{\mu}\right]} & \mbox{ , if } \Delta H(\sigma(\vec{x})\to\vec{x}') \geq 0 \end{cases}
\label{probability}
\end{equation}
Here $\Delta H(\sigma(\vec{x})\to\vec{x}')$ is the change in the Hamiltonian due to the attempted copy, and $\mu$ parametrizes the noise introduced by the active motion of cells. Immobilized cells are implemented by rejecting any copy from or into the space occupied by them ($\forall \vec{x} \in \Lambda_{\text{immob}} \lor \forall \vec{x}' \in \Lambda_{\text{immob}} : p(\sigma(\vec{x})\to \vec{x}')=0$). The Hamiltonian function consists of the sum of functions that describe distinct biological features:

\begin{equation}
H = H_v + H_{\chi} + H_a.
\label{Hamiltonian} 
\end{equation}
The first term is a volume constraint responsible for maintaining a controlled cell volume which we implemented in a novel way as:

\begin{equation}
H_v = \sum\limits_i \cfrac{\lambda_v(i)}{V^T(i)}\left( V(i) - V^T(i) \right)^2.
\label{Hvol} 
\end{equation}
Here $V(i)$ is the volume and $V^T(i)$ is the target volume of cell $i$ and $\lambda_v(i)$ describes cell compressibility. Our choice of the volume term differs from the originally proposed volume term \cite{Graner1992} as we divide $\lambda_v(i)$ with the target volume of the cell. Using the original formulation results in a synchronization of cell divisions in a compressed monolayer of cells. Using our novel formulation, the daughter cells after a cell division are less compressible than the mother, and thus a cell division does not trigger a wave of divisions by allowing the neighbors to expand more. Note that in a homogeneous population of cells (i.e., $\forall i\in\mathbb{Z}^{+}:V^T(i)=c$, with $c$ a constant) this formulation is equivalent to the original one.

The second term represents chemotaxis and is used to calculate a chemotactic bias for the change in the Hamiltonian as introduced by Savill and Hogeweg \cite{Savill1997}:

\begin{equation}
\Delta H_{\chi} (\sigma(\vec{x}) \to \vec{x}')= \sum\limits_{s\in \{O_2, g, l\}} \chi_{s} (s(\vec{x}') - s(\vec{x}))
\end{equation}
The function $s(\vec{x})$ represents the concentration of oxygen ($O_2(\vec{x})$) glucose ($g(\vec{x})$) or lactate ($l(\vec{x})$) at lattice site $\vec{x}$, and $\chi_s$ is the corresponding chemotactic coefficient. Diffusing chemicals are modeled with partial-differential equations, which are solved numerically on a grid matching with the grid of the CPM (see below).

For modeling cell-cell adhesion, the interactions are separated into surface-tension related adhesion and adhesion-molecule related adhesion \cite{Krieg:2008jd}, and are described by the last term of Eq.~\ref{Hamiltonian} as:

\begin{equation}
H_a = \sum\limits_{(\vec{x},\vec{x}')} (1-\delta (\vec{x}, \vec{x}')) J^{\text{eff}}(\sigma(\vec{x}),\sigma(\vec{x}'))
\end{equation}
where

\begin{equation}
\def\arraystretch{2.0}
\begin{array}{lcl}
J^{\text{eff}}(\sigma(\vec{x}),\sigma(\vec{x}')) & = & J^{\text{eff}}(i,i') \\
& = & J(\tau(i), \tau(i')) - \sum\limits_{a,b} k_{a,b} \times \min [\rho_a(i), \rho_b(i')]\\
\end{array}
\label{eq:jeff}
\end{equation}
and $(1-\delta (\vec{x}, \vec{x}'))=0$ if the sites $\vec{x}$, $\vec{x}' $ belong to the same cell, and $1$ otherwise. Above the simplified notation $i=\sigma(\vec{x})$ and $i'=\sigma(\vec{x}')$ is used for brevity. $J(\tau(i), \tau(i'))$ is the surface tension related adhesion coefficient between cell-types $\tau(i)$ and $\tau(i')$ present at positions $\vec{x}$ and $\vec{x}'$. As $J(\tau(i), \tau(i'))$ is typically positive, cells tend to minimize their surface area with other cells or the medium, making the adhesion term equivalent to surface tension \cite{Glazier1993}. Three cell types are used in the simulations: (stromal) cells, endothelial cells, and (cell-free) medium; these will be annotated as: $\tau(i) \in\{\mathrm{c}, \mathrm{EC}, \mathrm{m}\}$, respectively. The second term of $J^{\text{eff}}$ in Eq.~\ref{eq:jeff} describes the adhesion molecule related part of adhesion. The matrix $k_{a,b}$ defines the adhesion coefficient between adhesion molecule type $a$ and $b$, and $\rho_a(i)$ represents the density of molecule $a$ on cell $i$. Therefore, the maximum amount of adhesion related to adhesion molecules $a$ and $b$ between cells $i$ and $i'$ is determined by $ \min [\rho_a(i), \rho_b(i')]$. In our model, we chose the following specific adhesions between the existing cell types:

\begin{equation}
\def\arraystretch{2.0}
\begin{array}{lcl}
J^{\text{eff}}(i\mid_{\tau(i)=\mathrm{c}}, i'\mid_{\tau(i')=\mathrm{c}}) & = & 
J(\text{c},\text{c}) - k_{\text{CAM},\text{CAM}} \times \min[\rho_{\text{CAM}}(i), \rho_{\text{CAM}}(i')] \\
J^{\text{eff}}(i\mid_{\tau(i)=\mathrm{c}}, i'\mid_{\tau(i')=\mathrm{EC}}) & = & 
J(\text{c},\text{EC}) - k_{\text{CAM},\text{CAM}} \times \min[\rho_{\text{CAM}}(i), \rho_{\text{CAM}}(i')] \\
J^{\text{eff}}(i\mid_{\tau(i)=\mathrm{c}}, i'\mid_{\tau(i')=\mathrm{m}}) & = & 
J(\text{c},\text{m}) - k_{\text{MAM},\text{MAS}} \times \min[\rho_{\text{MAM}}(i), \rho_{\text{MAS}}(i')] \\
\end{array}
\end{equation}
Here CAM, MAM and MAS represent general ``cell adhesion molecules", ``matrix adhesion molecules" and ``matrix adhesion sites". CAM and MAM are expressed in cells but not the medium. MAS are expressed on medium (matrix) on which cells are allowed to adhere. In our simulations all the values of $k_{a,b}$ listed above have the value $1$, the rest of the $k_{a,b}$ matrix is set to $0$. 

The usual time measure in the CPM is the Monte Carlo Step (MCS). One MCS is defined as $N$ copy attempts, where $N$ is the number of lattice sites in the grid. This choice ensures that on average all sites are updated in every MCS, decoupling the system size and the number of copy-attempts needed to update the whole configuration.

\subsubsection*{Metabolism}
We implemented the following cellular metabolism system into the model. 

Cells use glucose and oxygen to produce internal energy that is used for the production of biomass and for supporting basal metabolism. In this section we describe how a cell produces energy from nutrients. 

Metabolites in the model are represented by the concentration fields of glucose, oxygen, and lactate, which diffuse and decay as:

\begin{equation}
\def\arraystretch{2.0}
\begin{array}{lclclcl}
\cfrac{dg(\vec{x},t)}{dt} & = & D_g \nabla^2 g(\vec{x},t) & - & \omega_g  g(\vec{x},t) & + & T_g(\sigma(\vec{x}),t) \\
\cfrac{dO_2(\vec{x},t)}{dt} & = & D_{O_2} \nabla^2 o(\vec{x},t) & - & \omega_{O_2}  O_2(\vec{x},t) & + & T_{O_2}(\sigma(\vec{x}),t) \\
\cfrac{dl(\vec{x},t)}{dt} & = & D_l \nabla^2 l(\vec{x},t) & - & \omega_l  l(\vec{x},t) & + & T_l(\sigma(\vec{x}),t) \\
\end{array}
\end{equation}
Here $D_s$ is the diffusion coefficient and $\omega_s$ is the decay constant of the respective species and $T_s$ describes the uptake or secretion of the chemical species according to boundary conditions (in ECs) and cellular metabolism. To ensure the desired trafficking of metabolites, we implemented cellular uptake of glucose and oxygen (sink) and secretion of lactate (source) while cells metabolize as described below. To simulate the diffusion process, the above fields are represented on the same lattice used for representing cells with periodic boundary conditions. 

Fields in functioning (active) blood vessel cells (ECs) are set as: $\forall \vec{x} \in \Lambda_{\text{EC}}: g(\vec{x})=1 \land O_2(\vec{x})=1 \land l(\vec{x})=0$, to represent the supply of glucose and oxygen and the removal of lactate. Non-functional (inactive) ECs do not alter nutrient levels. We implemented nutrient switching as follows: at every MCS each EC switches between active and inactive states with a switching probability $\mathcal{P}$. Setting $\mathcal{P}=0.5$ results in a frequent switching between active and inactive states providing a reliable steady nutrient supply. Lowering $\mathcal{P}$ increases the length of inactive and active states, leading to a more unreliable nutrient supply. This way $\mathcal{P}$ can alter the nutrient supply stability. A stable nutrient supply represents a normal vasculature, whereas tortuous neo-vascular vessels are blocked (inactivated) and unblocked (activated) with a higher variation. Here we relate the amount of tortuosity to the frequency of fluctuations in blood vessel function. Activation (unblocking) and inactivation (blocking) of ECs occur with the same probability, therefore simulations with different blocking frequencies $\mathcal{P}$ have the same average open and blocking times of blood vessels, and hence the same total amount of nutrients over time.

We implemented two different modes in which cells are able to use the nutrients, mimicking aerobic glycolysis and respiration of biological cells. Briefly, glucose is first degraded into two pyruvates which generates two ATPs. The pyruvate might be converted enzymatically to lactate (glycolysis) without further energy gain, or the cell might degrade it using oxydative phosphorylation (respiration) to gain $\alpha_r$ number of ATP molecules per glucose molecule ($\alpha_r > 2$).

In the model, cells take up glucose through GLUT transporters using a Michaelis-Menten-like dynamics with cell- and time-specific constants: for cell $i$ at time $t$ the maximum rate is denoted as $V^m(i,t)$ and the Michaelis-Menten constant is denoted as $K(i,t)$. The uptake of glucose from the environment at position $\vec{x}$ per unit time is given by:

\begin{equation}
u_g(\vec{x}, t) = \left(\frac{N_G(i,t)}{S(i),t)}\right) \cdot \left(\frac{V^m(i,t) g(\vec{x},t)}{K(i,t)+g(\vec{x},t)}\right)
\end{equation}
Again, $i=\sigma(\vec{x})$. Here $g(x,t)$ is the extracellular glucose concentration at point $\vec{x}$ and time $t$, $N_G(i,t)$ is the number of glucose transporters (in arbitrary units) on the surface of cell $i$, and $S(i,t)$ is the cell's surface (assuming homogeneous GLUT distribution). We implemented the total amount of glucose taken up by cell $i$ in each MCS as the sum of uptakes over the cell surface: 

\begin{equation}
\def\arraystretch{2.0}
\begin{array}{lcl}
U_g(i,t) & = & \sum\limits_{\vec{x}' \in S(i,t)} u_g(\vec{x}', t) \\
& = & N_{G}(i,t) \cfrac{V^m(i,t) g_i(t)}{K(i,t)+g_i(t)} \\
\end{array}
\label{eq:glucoseUptake}
\end{equation}
where $g_i(t) = \sum_{\vec{x}' \in S(i,t)} g(\vec{x}',t)$ is the total amount of glucose at the surface of cell $i$ ($i=\sigma(\vec{x})$). 

A part of the glucose, denoted by $h(i,t)$ for cell $i$ at time $t$, is used only in glycolysis to produce 2 lactate molecules and 2 ATP molecules per glucose molecule ($0\leq h(i,t) \leq 1$). The rest of the glucose ($1-h(i,t)$) is directed to oxydative phosphorylation after glycolysis to produce $\alpha_r$ ATP molecules per 1 glucose molecule in total. Respiration requires 6 oxygen molecules per 1 glucose molecule, and produces reactive oxygen species (ROS, $\zeta(i,t)$ in cell $i$ at time $t$) as a side-product. Therefore in our implementation the uptake of oxygen is determined by the uptake of glucose and the ratio $h(i,t)$:

\begin{equation}
u_{O_2}(\vec{x},t)) = 6 (1-h(i,t)) u_g(\vec{x},t)
\end{equation}
at every point $\vec{x}$ of the surface of cell $i=\sigma(\vec{x})$ at time $t$. Here we use the experimental results of Subczynski and colleagues \cite{Subczynski1992}, who showed that oxygen permeability cannot represent a constraint to cellular respiration. Thus, we assume that the internal and external oxygen concentrations equilibrate much faster than the metabolic processes that use internal oxygen. We implemented the energy gain of cell $i$ at time $t$ as the sum of energy produced by fermentation and respiration, minus the cost of basal metabolism:

\begin{equation}
\Delta E(i,t) = \bigg( 2 h(i,t) + \alpha_r (1-h(i,t))\bigg)U_g(i,t) - E_m \Delta t
\label{eq:energy_production} 
\end{equation}
where $E_m \Delta t$ represents the basal metabolic cost for the cell during a time interval $\Delta t$, and is considered to be a fixed constant in the model. Therefore, the variable $h(i,t)$ describes the state of hypoxia in the cell, similar to the transcription factor HIF1-$\alpha$. If $h(i,t)=0$, the cell is using all the glucose in the oxidative phosphorylation cycle. If $h(i,t)=1$, respiration is completely blocked and the cell changes into a state of complete fermentation in which all glucose is turned over into lactate. In order to compensate for the reduced efficiency in energy production in a controlled manner, when the cell is fermenting, $h(i,t)=1$, we assumed the number of glucose transporters is upregulated as well (consistent with experimental observations \cite{Jones2009}):

\begin{equation}
N_G(i, t) = N_0(i, t) \cdot \cfrac{\alpha_r - 2}{\alpha_r-(\alpha_r-2)h(i, t)}
\label{eq:nglut}
\end{equation}
Here $N_0(i,t)$ is the factor in the GLUT concentration that is independent of the mode of cell metabolism, and sets the intrinsic growth signal of cell $i$ at time $t$. This choice of function describes a controlled compensation for the hypoxia, since the energy gained by cell $i$ in a MCS is independent of the mode of metabolism, $h(i,t)$. Substituting from Eqs.~\ref{eq:energy_production}, \ref{eq:glucoseUptake}, and \ref{eq:nglut}:

\begin{equation}
\def\arraystretch{2.5}
\begin{array}{lll}
\cfrac{\partial\Delta E(i,t)}{\partial h(i,t)} 
& = \cfrac{\partial}{\partial h(i,t)}\Bigg[ & \bigg((2-\alpha_r)h(i,t) + \alpha_r\bigg) \cdot \cfrac{N_0(i, t) (\alpha_r - 2)}
{\alpha_r-(\alpha_r-2)h(i, t)}\\
&& \cdot \cfrac{V^m(i,t)g_i(t)}{K(i,t)+g_i(t)} + E_m\Delta t \Bigg] \\

& = \cfrac{\partial}{\partial h(i,t)}\Bigg[ & \cfrac{N_0(i, t) (\alpha_r - 2)V^m(i,t)g_i(t)}{K(i,t)+g_i(t)} + E_m\Delta t\Bigg] = 0
\end{array}
\label{eq:energy_dependence}
\end{equation}

The amount of $N_G(i,t)$ is calculated based on $g_i(t)$, the average available glucose concentration over the whole cell surface (before consumption), as the amount of $N_G$ is controlled by the cell as a whole. We set a maximum surface density limit on the GLUT molecules, by limiting parameter $N_0(i,t)$ to a fixed value.

The hypoxia inducible factor, HIF1-$\alpha$, is induced if the level of oxygen is low in the cell: under normal circumstances PHDs (prolyl hydroxylase) combine with oxygen and inactivate HIF1-$\alpha$ \cite{Denko2008}. When oxygen is low, the degradation of HIF1-$\alpha$ is slower, allowing an increase in HIF1-$\alpha$ levels, leading to the reduction of respiration and therefore oxygen use. Furthermore, HIF1-$\alpha$ is stabilised by the reactive oxygen species (ROS) produced in the mitochondrial respiratory cycle in biological cells, creating a negative feedback to downregulate respiration if too much ROS is present \cite{Denko2008}. To model these dependencies, we implement $h(i,t)$ as a function of intracellular ROS ($\zeta(i,t)$) and average oxygen level in the cell using a combination of Hill functions:

\begin{equation}
\def\arraystretch{2.5}
\begin{array}{lcl}
h(i,t+1) & = & h(\zeta(i,t), \langle O_2\rangle(i,t)) \\
& = & 1 -\left[ \cfrac{\langle O_2\rangle(i,t)^{n_h(i,t)}}{\kappa_h(i,t)^{n_h(i,t)}+ \langle O_2\rangle(i,t)^{n_h(i,t)}} \right] \\
&& \cdot \left[ 1 - \cfrac{\zeta(i,t)^{n_\zeta (i,t)} }{\kappa_\zeta(i,t)^{n_\zeta (i,t)}+\zeta(i,t)^{n_\zeta (i,t)}}\right]\\
\end{array}
\label{eq:hif} 
\end{equation}
Here $\kappa(i,t)$ and $n(i,t)$ are parameters describing the threshold and sensitivity of the switch to the levels of ROS ($\zeta(i,t)$) and average oxygen level ($\langle O_2\rangle(i,t)$) in cell $i$ at time $t$. Note that the value of $h(i,t)$ in the cell is calculated from values of ROS and oxygen within the cell at the previous time step $t$. This time delay is natural, as the signaling is not expected to instantly alter the biochemical state of the cell. Finally, the level of ROS is increased by the amount of glucose processed in mitochondrial respiration in the previous time step, and the previous values of ROS in the cell:

\begin{equation}
\zeta(i,t+1) = (1-h(i,t))U_g(i,t) - \omega_\zeta \zeta(i,t)
\end{equation}
where $\omega_\zeta$ is the decay constant of the ROS inside the cell.

\subsubsection*{Cell growth}

In every time step, the cell is allowed to use its energy to produce biomass, by increasing its target volume $V^T(i,t)$ as:

\begin{equation}
V^T(i,t+\Delta t)=V^T(i,t) + \cfrac{\partial V^T(i,t)}{\partial t}\Delta t 
\end{equation}
with:

\begin{equation}
\cfrac{\partial V^T(i,t)}{\partial t} = \alpha \cdot (E(i,t)- E_{m}) - \beta \cfrac{(V(i,t)-V^T(i,t))^3}{|V(i,t)-V^T(i,t)|} - \gamma \Theta(\zeta(i,t) - \theta_R(i,t))
\label{eq:cellGrowth} 
\end{equation}
where $\alpha$ is a growth rate parameter, describing the efficiency of anabolism: biomass produced per unit of energy per unit of time, and $E_{m}$ is the energy lost for the basal metabolism per time step ($\Delta t$). The second term expresses a tendency of a model cell to reduce its intracellular density fluctuations: If a cell is compressed, its growth is decreased. The third term expresses the effect of ROS ($\zeta$) on cell growth. High levels of ROS are toxic for the cell (interferes with protein folding), forcing the model cell to reduce in size. The function $\Theta$ is the Heaviside step-function and $\theta_R(i,t)$ is a threshold for ROS levels that a cell can tolerate without damage. If cells are not allowed to grow, for example due to contact inhibition of growth, then the target volume is not increased, but is allowed to decrease. 

Irrespective of whether a cell grows or not, the energy of the cell is depleted leaving the cell with no internal energy reserve. If the produced energy in a time step is smaller than the metabolic need of the cell ($E(i,t) - E_m$), the cell's target volume will be reduced by $\alpha \cdot (E(i,t)- E_{m})$ and its energy is set to zero. This way we include catabolism in our model cells, which allows them to use their target volume as a type of internal energy storage. 

Cells divide if their size ($V(i,t)$) is around division size $V_D(i,t)$. The probability of allowing cell division is defined by a sigmoid function:

\begin{equation}
P_{\mbox{size}}(\mbox{div})=
\begin{cases} 0 & \text{, if } V(i,t) < V_D(i,t) - 3w \\
\cfrac{V(i,t)-V_D(i,t)}{2 \sqrt{w^2 + (V(i,t)-V_D(i,t))^2}} + \cfrac{1}{2} & \text{, if } V_D(i,t) - 3w \leq V(i,t) < V_D(i,t) + 3w \\
1 & \text{, if } V_D(i,t) + 3w \leq V(i,t) \\
\end{cases}
\end{equation}
Here $w$ is a parameter setting the width of the fuzzy threshold, and is set to the 5\% of the division size: $w=0.05 V_D(i,t)$. Upon division, the mother cell divides, and its biomass ($V^T(i,t)$) is shared equally between the two daughter cells. 

To maintain cell turnover, cells are selected for apoptosis at a constant probability (0.1\% per cell per MCS). When a cell is selected for apoptosis, its target volume is decreased with a constant rate until it reaches zero. At that point the cell is removed from the simulation and its volume is converted into extracellular space.

\subsubsection*{Evolution in the model}

We implemented evolution in the model as follows. To mimic plasticity, a set of cellular parameters are tested independently for mutation in daughter cells. Such a parameter $p$ is allowed to change with a probability (mutation rate $\mu_p$). The new value for the parameter is drawn from a normal distribution centered at the previous value, and with a standard deviation determined by the characteristic step size parameter $\sigma_p$ of the mutation for each parameter: $p' = p + \xi(\sigma_p)$, where $\xi(\sigma_p)$ is drawn from the normal distribution $\mathcal{N}(0,\sigma_p^2)$. The parameter values are bound in within a pre-set parameter range, with reflective boundaries (see Fig.~\ref{fig_modelSetup}b). Constructed this way, the cells perform a random walk in the parameter space through mutations as they progress from generation to generation. 

The cellular parameters that are allowed to undergo such mutation are controlling the physical properties of the cells (compressibility through $\lambda_{v}$, division size $V_D$), adhesion properties ($\rho_{\text{CAM}}(\text{c}), \rho_{\text{MAM}}(\text{c})$), chemotaxis sensitivities ($\chi_{g}, \chi_{O_2}, \chi_{l}$), growth (through $N_0$), and metabolism (sensitivity thresholds $\kappa_h, \kappa_\zeta$). Values for these parameters are summarized in Table \ref{table}.
\vspace{0.3cm}

\subsection*{Initial conditions}

Cells are initialized in a monolayer with an initial volume and target volume of 25 lattice sites on a lattice of $200\times200$ and four endothelial cells (Fig.~\ref{fig_modelSetup}a). In the initial regime of the simulations, nutrients are allowed to diffuse into the system to obtain a natural distribution resulting in high glucose and oxygen and low lactate levels (Fig.~\ref{fig_modelSetup}b). During this equilibrating time cells are allowed to metabolize, but cannot grow, shrink, move, or mutate. This state represents a stable, homeostatic tissue with sufficient nutrient availability. When the temporal changes in diffusing nutrients is less than 5\% in a time interval of 100 MCS, the initial regime is closed, cells are released and the simulation is started. Nutrient fields at the beginning of the initial regime (Fig.~\ref{fig_modelSetup}b) are initialized using pre-generated concentration distributions to expedite equilibration. These initial concentration fields are generated by simulating a cell population in the initial regime starting with zero concentrations: $\forall\vec{x}\in \Lambda: g(\vec{x}, t=0) = 0$, $O_2(\vec{x},t=0)=0$, $l(\vec{x}, t=0)=0$. The concentration fields are saved when the total concentration levels remain within 5\% over a 100 MCS iteration period: $\sum_{\vec{x}} (s(\vec{x},t=t_{\text{save}}) - s(\vec{x},t=t_{\text{save}}-100 \text{MCS})) < 0.05 \sum_{\vec{x}} s(\vec{x},t=t_{\text{save}}-100 \text{MCS})$ for $s\in\{g,O_2\}$. Note that no lactate is present as all cells are respiratory in the initial regime. All mutating parameters in all cells are initialized with the same value, after which all cells undergo a mutation attempt to provide an initial heterogeneity to the population. 

Adhesion parameters are set to $J(\text{c},\text{c})=J(\text{c},\text{EC})=J_{\text{max}}$, and $J(\text{c},\text{m})=J_{\text{max}} /2 $ for the three region (``cell'') types: stromal cells (c), endothelial cells (EC), and cell-free areas (m). To allow the full range of interactions for cells, we fix the adhesion molecule density values of the extracellular region and the endothelial cells as:

\begin{equation}
\begin{array}{lclcl}
k_{\text{CAM},\text{CAM}} \times \rho_{\text{CAM}}(i\mid_{\tau(i):\text{EC}}) & = & J(\text{c},\text{EC}) & = & J_{\text{max}} \\
k_{\text{MAM},\text{MAS}} \times \rho_{\text{MAS}}(i\mid_{\tau(i):\text{m}}) & = & J(\text{c},\text{m}) & = & J_{\text{max}} /2 
\end{array}
\end{equation}
With this choice the effective $J^{\text{eff}}$ values are allowed to change between $0$ and $J_{\text{max}}$ for $J^{\text{eff}}(\text{c},\text{c})$, and $J^{\text{eff}}(\text{c},\text{EC})$. For interaction with the medium, $J^{\text{eff}}(\text{c},\text{m})$ is allowed to change between $0$ and $J_{\text{max}} /2$.

Diffusion parameters for glucose, oxygen, and lactate were set to $D_g=10^{-9}m^2/s$ and $D_{O_2}=D_l=10^{-11}m^2/s$ following the approximation of Jiang and coworkers \cite{Jiang2005}, and decay is neglected for all of these chemical species ($\omega_s=0, s\in\{g, O_2, l\}$). Decay rate of ROS is set to a constant $\omega_\zeta=0.1$ per MCS. The amount of ATP produced from 1 glucose molecule through respiration was chosen as $\alpha_r=38$. This approximation is a theoretical upper limit for the process, in reality this number is expected to be lower. However, due to the compensation with the number of glucose transporters (see Eq. \ref{eq:nglut}) the exact value of this parameter is not expected to change the behavior of the system.

\subsection*{Parameter analysis}

To analyse the population behavior over the simulations we analyzed the evolving parameters in the following way. For every parameter $p(i,t)$ of cell $i$ at time $t$ we calculated a normalized parameter as $p_n(i,t) = [p(i,t) - p(i,t=0)]/\sigma_p$ where $\sigma_p$ is the characteristic step size of the parameter. This way the normalized parameters reflect their distance from their origin in terms of mutational step-size. The distribution of cell phenotypes in this space was analyzed at time $t$ by finding the principal components of the 10-dimensional set of $p_n(i,t)$ points for all $i$ and $p$ using singular value decomposition from scientific python (SciPy). The axes corresponding to the first three largest eigenvalues were selected as the main components of the cloud. 

Clustering in the normalized phenotype space was performed using hierarchical clustering of SciPy with the Ward method and Euclidean distances. To distinguish clusters we established a cutoff cophenetic distance of 100 manually by evaluating a set of selected dendograms and distribution of clusters plotted on the first three main axes. Displacement of the clusters is calculated as the (10D) Euclidean distance of the center of mass of the cluster from its initial point of origin. Spread of clusters was calculated as the mean distance of points in the cluster from its center of mass. Density of clusters was calculated by dividing the spread by the number of points in the cluster.

\section*{Supporting Information}

\paragraph*{S1 File.}
\label{S1-File}
{\bf Script and code for running simulations of the model using the CompuCell3D framework (ZIP).} Simulations were run using a CompuCell3D installation available from the Indiana University Bloomington and the Biocomplexity 
Institute (www.compucell3d.org) \cite{Swat2012}. Parameters for the simulations are set in XML scripts ({\em simulation.xml}). Simulations require initial configuration files for cells and diffusible concentration fields ({\em PIFtemplate.pif, FieldGLUtemplate.txt, FieldOXYtemplate.txt}) specified in the XML file, and a set of customized plugins (called 'steppables' and 'plugins') to be installed in the CompuCell3D `DeveloperZone'. Simulation outputs are a set of files for each sampling time point that contain cell-based information (values of mutating parameters and other cell-level indicators), concentration field data, and cell configuration data.

\section*{Acknowledgments}
The authors thank Indiana University Bloomington and the Biocomplexity Institute for providing the CompuCell3D modeling environment (www.compucell3d.org) \cite{Swat2012}. This work was carried out on the Dutch national e-infrastructure with the support of SURF Cooperative (www.surfsara.nl). 

This work was cofinanced by the Netherlands Consortium for Systems Biology (NCSB), which is part of the Netherlands Genomics Initiative/Netherlands Organization for Scientific Research (NWO). The work is also part of the research programme “Innovational Research Incentives Scheme Vidi Cross-divisional 2010 ALW” with project number 864.10.009, which is (partly) financed by the Netherlands Organisation for Scientific Research (NWO). The funders had no role in study design, data collection and analysis, decision to publish, or preparation of the manuscript.



\end{document}